\documentclass[prb,twocolumn,longbibliography]{revtex4-2}
\usepackage[dvips]{graphicx}
\usepackage{float, color,array, multirow}

\begin{document}

\title{Ground state features and spectral properties of large polaron liquids from low to high charge densities}

\author{C. A. Perroni$^{1,2}$, G. De Filippis$^{1,2}$, and V. Cataudella$^{1,2}$}

\date{\today}

\affiliation{${}^{1}$ Physics Department E.  Pancini, Universit\`a
degli Studi  di  Napoli  Federico  II, Complesso Universitario
Monte  S. Angelo,  Via  Cintia,
I-80126 Napoli, Italy \\
${}^{2}$CNR-SPIN  c/o Universit\`a degli Studi  di Napoli Federico
II, Complesso Universitario Monte  S. Angelo,  Via  Cintia,
I-80126 Napoli, Italy}

%\affiliation{${}^{1}$Department of Earth and Space Science, Graduate School of Science, Osaka University, Osaka 560-0043, Japan \\
%${}^{2}$Toyota Physical and Chemical Research Institute, Aichi,
%480-1118, Japan }

\begin{abstract}
A new variational approach is proposed at zero temperature for a
finite density of charge carriers in order to study ground state
features of the Fr\"ohlich model including electron-electron and
electron-phonon interactions. Within the intermediate
electron-phonon coupling regime characteristic of large polarons,
the approach takes into account on the same footing polaron
formation and polaron-polaron correlations which play a relevant
role going from low to high charge densities. Including
fluctuations on top of the variational approach, the electronic
spectral function is calculated from the weak to the intermediate
electron-phonon coupling regime finding a peak-dip-hump line
shape. The spectra are characterized by a transfer of spectral
weight from the incoherent hump to the coherent peak with
decreasing the electron-phonon coupling constant or with
increasing the particle density. Three different density regimes
stem out: the first, at low densities, where the features of a
single large polaron with a substantial incoherent spectral weight
are not modified by charge carrier interactions; a second one, at
intermediate densities, where the polaronic liquid shows a rapid
crossover from incoherent to coherent dynamics; the third one, at
high densities, where screening effects are so prominent that the
system presents a conventional metallic phase. The results
obtained in the low to intermediate density regime turn out to be
relevant for the interpretation of recent tunneling and
photoemission experiments in $SrTiO_3$-based systems.
\end{abstract}

\maketitle
\section{introduction}

The polaron is a fermionic quasiparticle which takes into account
the interaction of an electron with lattice vibrations in a solid
\cite{alex1,alex2,mahan}. This concept has been originally used in
polar semiconductors to indicate that the polarization cloud
follows the electron in its motion. Indeed, a first classification
of the polaron is based on the size of the phonon cloud. If the
electron-phonon coupling is not very strong, the phonon cloud
accompanying the electron extends over lengths larger than the
lattice parameter of the solid, therefore the corresponding
polaron is termed large. Large polarons are itinerant
quasi-particles whose dynamics affects the spectral, transport and
optical properties of solids.

In the last years $SrTiO_3$-based (STO) systems  have become one
of the main research areas of the condensed matter community
\cite{caviglia,levy}. Not only the three-dimensional (3D) STO bulk
but also the two-dimensional (2D) STO surface and quasi-2D
heterostuctures, such as those between STO and $LaAlO_3$ (LAO),
have been much studied showing many interesting properties among
which superconducting phases strongly tunable by chemical doping
(in the bulk) or by application of a gate potential (in the
heterostructures). In particular, at odds with simple metallic
systems, superconducting states occur at quite low carrier
densities \cite{lin1,lin2} suggesting the presence of large
pairing potentials and novel features of the normal state. In
order to interpret the bulk data from angle-resolved photoemission
spectroscopy (ARPES) \cite{aiura,takizawa,chang,Meevasana1}, it
has been suggested that a substantial interaction between
electrons and lattice distortions plays a significant role. The
relevance of large polaron quasi-particles \cite{strocov} has been
confirmed by the experimental spectral properties not only of the
3D bulk \cite{swartz}, but also the 2D surface
\cite{devereaux,chen} and LAO/STO heterostructure
\cite{cancellieri}. In particular, the large polaron formation is
promoted by the sizable coupling of the electron with a
well-defined high frequency longitudinal optical mode
\cite{swartz,devereaux,cancellieri}. In addition to tunneling and
photoemission spectra, the optical properties
\cite{Mechelen,dubroka} and inelastic x-ray scattering
measurements \cite{Salluzzo} have shown that the charge carriers
undergo a crossover from a polaronic liquid to a Fermi liquid
regime with increasing density. Nowadays the role of large
polarons is widely recognized in STO-based materials.

The Fr\"ohlich model has been frequently used to simulate the
large polaron formation when the relevant coupling is between
electrons and longitudinal optical phonons in polar materials
\cite{alex1,alex2,mahan,cataud}. In STO-based systems,
experimental data have been interpreted within the  Fr\"ohlich
model \cite{strocov} suggesting that the electron-phonon
interaction is not perturbative, but in the intermediate coupling
regime. For the single Fr\"ohlich polaron, the different coupling
regimes of the electron-phonon interaction have been investigated
by several variational approaches \cite{alex2,cataud,giulio1}
which provide also the starting point for excited state properties
\cite{giulio2}. A simple variational approach  is that based on
the  Lee-Low-Pines (LLP) canonical transformation which is quite
accurate in the weak to intermediate electron-phonon coupling
regime \cite{LLP,chatterjee}. All the results of these variational
approaches for the single polaron have been checked by many
numerical methods, among which that based on the diagrammatic
quantum Monte Carlo (DQMC) \cite{strocov,andrei1} is one of the
most accurate for all the electron-phonon couplings.

Theory for many polaron systems is feasible for weak
electron-phonon couplings \cite{mahan,swartz,devereaux}, but it is
quite challenging for non-perturbative regimes
\cite{strocov,andrei2,giulio,perroni2}. Considering only the
effects of the electron-phonon interaction, a recent theoretical
study has shown that the crossover from polarons to Fermi liquids
in transition metal oxides, among which titanates, occurs when the
frequency of plasma oscillations exceeds that of longitudinal
optical phonons \cite{verdi}. For STO-based systems, the optical
response has been calculated starting from the Fr\"ohlich model
including Coulomb electron-electron interactions
\cite{Devreese,Klimin}. However, many features of the full
Fr\"ohlich model, such as the spectral properties, are not fully
understood  in the intermediate electron-phonon coupling regime.
One possibility to approach this problem is to generalize the
variational approaches for the single polaron to the many particle
case \cite{alex1}. Indeed, the LLP canonical transformation has
been performed in second quantization to treat variationally the
electron-phonon interaction for a finite density of charge
carriers \cite{Lemmens}. One drawback of these approaches is that,
after the canonical transformation to the polaron configuration
space, polaron-polaron interactions have been treated only at
Hartree-Fock level. It is highly desirable to treat the electron
liquid with polaronic effects beyond the mean-field theory
\cite{bassani}.

In this paper, in order to analyze the ground state properties of
the Fr\"ohlich model in the intermediate electron-phonon coupling
regime, we explicitly include polaron-polaron correlations after
the variational many-particle LLP canonical transformation.
Actually, polaron-polaron interactions are taken into account
through a variational Slater-Jastrow term in the many-body
wave-function which, therefore,  includes the suppression of long
wavelength density fuctuations \cite{Gaskell}. In fact, the
treatment of charge correlations is at the same level of the
many-body approach known as Random Phase Approximation (RPA)
\cite{mahan}. Many quantities, such as the static structure factor
and the polaronic band shift, have been evaluated  pointing out
that polaron-polaron correlations represent very relevant effects
with increasing particle density.

The electronic spectral function is calculated from the weak to
the intermediate electron-phonon coupling regime for different
carrier concentrations by including fluctuations on top of the
variational approach. The large polaron spectra are characterized
by a peak-dip-hump line shape. For the single polaron, in the
intermediate electron-phonon coupling regime, the hump has a
relevant spectral weight and it consists of several phonon
satellites  in good agreement with numerical approaches. Screening
promotes a transfer of spectral weight from the incoherent hump to
the coherent peak with increasing the particle density. In
agreement with recent tunneling and photoemission experiments in
STO-based systems \cite{swartz,devereaux,cancellieri}, we identify
three different density regimes: the low density one, where the
spectra bear a strong resemblance to those of the single large
polaron; the intermediate density one, where the crossover from
incoherent to coherent dynamics is quite rapid; the high density
one, where the system behaves as a conventional metal. It turns
out that the role of density can be roughly understood as an
effect leading to the reduction of the effective electron-phonon
coupling constant. However, for intermediate electron-phonon
couplings, the density evolution of the spectral properties can
not be ascribed only to many-body screening, but also to polaron
features ranging from the antiadiabatic to the adiabatic regime.
Our results are fully consistent with experimental findings in
STO-based systems clarifying the role of the electron-phonon
coupling in the low to intermediate density regime.

The paper is organized as follows. In section II the model and the
variational approach are reviewed; in section III the spectral
properties are discussed; in section IV conclusions and
discussions. We present additional details about polaron-phonon
couplings in Appendix A, and polaronic spectral features in
Appendix B.

\section{The model and the variational approach}
In this paper, the Fr\"ohlich model \cite{mahan,LLP} is studied
focusing on the normal state of $N$ polarons at zero temperature.
The starting point of the model is the jellium for interacting
electrons. In addition to the Coulomb electron-electron
interaction, the model takes into account the coupling between
electrons and longitudinal optical phonons. The long-range
electron-phonon interaction is derived under the assumption that
the medium is a polarizable continuum with partially ionic
character. The  Fr\"ohlich model has been extensively used for the
description of doped polar semiconductors \cite{mahan}.

The Hamiltonian $H$ of the Fr\"ohlich model in second quantization is the following:
\begin{equation}
H= H_{el}^{(0)} + H_{ph}^{(0)} +H_{el-el}^{\infty} + H_{el-ph},
 \label{0r}
\end{equation}
where the first term $H_{el}^{(0)}$, defined as
\begin{equation}
H_{el}^{(0)}= \sum_{{\bf k},\sigma}  \frac{\hbar^2 k^2}{2m}  c^{\dagger}_{{\bf k},\sigma} c_{{\bf k},\sigma},
\label{0r1}
\end{equation}
describes the conduction band electrons of wave-vector ${\bf k}$ ($k=|{\bf k }|$ its modulus), mass $m$ and
spin $\sigma$, with $ c^{\dagger}_{{\bf k},\sigma} \left( c_{{\bf k },\sigma} \right)$ the related creation (annihilation) operator,
while the second term $H_{ph}^{(0)}$, defined as
\begin{equation}
H_{ph}^{(0)}= \hbar \omega_{LO} \sum_{ {\bf q} } a^{\dagger}_{{\bf q}} a_{{\bf q}},
\label{0r2}
\end{equation}
characterizes the energy of free longitudinal optical phonons of wave-vector ${\bf q}$ and  angular frequency $\omega_{LO}$, with
$ a^{\dagger}_{{\bf q}} \left( a_{{\bf q}} \right)$ related  creation (annihilation) phonon operator.

In Eq. (\ref{0r}), the electron-electron interaction is provided by the following Hamiltonian $H_{el-el}^{\infty}$
\begin{equation}
H_{el-el}^{\infty}= \frac{1}{2V} \sum_{{\bf q}}
V_q^{\infty} \left( \rho_{\bf q} \rho^{\dagger}_{\bf q} - N \right),
\label{1r}
\end{equation}
where  $V$ is the volume of the system, $\rho_q$ is the density operator
\begin{equation}
\rho_{\bf q} =  \sum_{{\bf k},\sigma}
c^{\dagger}_{{\bf k}+{\bf q},\sigma}   c_{{\bf k},\sigma},
\label{1rter}
\end{equation}
and $V_q^{\infty}$ is the Coulomb potential
\begin{equation}
V_q^{\infty}=  \frac{4 \pi e^2}{\epsilon_{\infty} q^2},
\label{1rbis}
\end{equation}
with $e$ the modulus of the electron charge, $\epsilon_{\infty}$
the dielectric function at frequencies higher than those of the
optical modes, $\hbar q$ the modulus of the momentum exchanged by
the electrons. Actually, the dielectric constant   $
\epsilon_{\infty}$ takes into account electronic excitations
across the semiconductor gap, which are therefore at high
energies. Indeed, these electronic excitations provide a constant
contribution on the low energy scale of vibrational modes and
conduction electrons close to the Fermi energy.

In Eq. (\ref{0r}), the electron-phonon interaction is given by the following Hamiltonian $ H_{el-ph}$
\begin{equation}
H_{el-ph} =\frac{1}{\sqrt{V}}
\sum_{{\bf q}} M_q \rho_{{\bf q}} \left( a_{{\bf q}}+a^{\dagger}_{-{\bf q}}  \right),
\label{2r}
\end{equation}
where the electron-phonon matrix element $M_q$ is
\begin{equation}
M_q =  \hbar \omega_{LO} \frac{ \sqrt{4 \pi \alpha R_p} }{q}.
\label{2rbis}
\end{equation}
In Eq. (\ref{2rbis}), the dimensionless electron-phonon coupling constant  $\alpha$, defined as
\begin{equation}
\alpha =  \frac{e^2}{2 R_p \hbar \omega_{LO}} \left( \frac{1}{\epsilon_{\infty}}-\frac{1}{\epsilon_0} \right),
\label{3r}
\end{equation}
is determined not only by $\epsilon_{\infty}$, but also by the
static dielectric constant $\epsilon_0$, therefore it depends on
the polarizability of the system. Moreover, $R_p$ is the polaron
radius defined as
\begin{equation}
R_p =  \sqrt{ \frac{\hbar}{ 2 m \omega_{LO} } }.
\label{4r}
\end{equation}

The parameters of the many-body Hamiltonian (\ref{0r}) are the
electron mass $m$, the phonon angular frequency $\omega_{LO}$, the
dielectric constants $\epsilon_{\infty}$ and $\epsilon_0$. Another
important quantity in this paper is the particle density $n=N/V$,
which determines the Fermi wave-vector $k_F$. In the case of
STO-based systems, the following values are assumed
\cite{levy,Devreese,ruhman}: $m \simeq 2 m_0$, with $m_0$ electron
rest mass, $\omega_{LO} \simeq 2.42 \cdot 10^{13} s^{-1}$
(corresponding to $\hbar \omega_{LO} \simeq 100$ meV),
$\epsilon_{\infty} \simeq 5.1$, and $\epsilon_{0} \simeq 2 \cdot
10^{4}$. This high frequency phonon mode  is the most coupled to
the electrons and it is clearly discernible in experimental
measurements \cite{chen,devereaux,cancellieri}.  We remark that a
model with a single electronic band does not represent a
limitation for the analysis of STO-based systems, since, in the
ARPES setup, it is possible to use polarized light in order to
investigate selected electronic bands. For example, using
s-polarized light \cite{devereaux}, the measurements can resolve a
single $d_{xy}$ band at low density.

By using the Hamiltonian parameters and Eq. (\ref{3r}), one gets
the electron-phonon coupling constant $\alpha \simeq3.37$, that is
STO-based systems are well within the intermediate electron-phonon
coupling regime. This non-perturbative coupling regime is
notoriously difficult to analyze in systems, like STO-based
compounds, whose static and dynamic properties are sensitive to
the variations of particle density. Indeed, an important feature
of STO-based systems is the possibility to tune the particle
density over several orders of magnitude. In the next subsection,
we expose our new variational approach which takes into account
both the polaron formation and the effects of polaron-polaron
interactions from weak to intermediate electron-phonon coupling
regime ranging from low to high densities.

\subsection{Variational approach}

Following a proposed variational scheme valid for a finite number
$N$ of electrons \cite{Lemmens},  the  LLP canonical
transformation \cite{LLP}  is performed in second quantization to
treat variationally the electron-phonon interaction  up to the
intermediate coupling regime ($\alpha < 6$). The variational
unitary transformation $U$ is
\begin{equation}
U=\exp \left[ \frac{1}{\sqrt{V}} \sum_{{\bf q}}  f_q  \rho_{{\bf q}} \left( a_{{\bf q}}-a^{\dagger}_{- {\bf q} } \right) \right],
\label{5r}
\end{equation}
where $f_q$ is a real variational function, which  provides the
phonon distribution function induced by the electron dynamics.
Actually, the function  $f_q$ shifts the position of the
vibrational modes quantifying the strength of the coupling between
electron and lattice displacement, hence it measures the degree of
polaronic effect.  We point out that, even in the intermediate
electron-phonon coupling regime, the interactions at zero
temperature are not able to localize the electron, that this way
behaves as an itinerant large polaron \cite{alex1,alex2,cataud}.

The transformed Hamiltonian $\tilde{H}= U^{-1} H U $ describes the interaction between large polarons and shifted vibrational modes being
\begin{equation}
\tilde{H}= H_{ph}^{(0)}+H_{pol} +  H_{pol-ph} + H_{pol-2ph},
\label{6r}
\end{equation}
where $H_{ph}^{(0)}$ is the free phonon Hamiltonian equal to Eq.
(\ref{0r2}), and $H_{pol}$ describes many interacting large
polarons:
\begin{equation}
H_{pol}= H_{pol}^{(0)} + H_{pol-pol}. \label{18r}
\end{equation}
In Eq. (\ref{18r}), the free polaron Hamiltonian $H_{pol}^{(0)}$
has the same form as the kinetic energy in Eq. (\ref{0r1}), but
the quadratic term in the momentum is replaced by the polaronic
band $\epsilon_k$
\begin{equation}
\epsilon_k =   \frac{\hbar^2 k^2}{2m} + \eta,
\label{7r}
\end{equation}
with the polaronic band shift $\eta$ given by
\begin{equation}
\eta = \frac{1}{V} \sum_{{\bf q}} \left( \hbar \omega_{LO}+  \frac{\hbar^2 q^2}{2 m} \right) f^2_q - \frac{2}{V} \sum_{{\bf q}} M_q f_q.
\label{8r}
\end{equation}
In Eq. (\ref{18r}), the polaron-polaron interaction term
$H_{pol-pol} $ has the same form as Eq. (\ref{1r}), with
$V_q^{\infty}$ replaced by the following effective potential
$V_q^{eff}$:
\begin{equation}
V_q^{eff}=V_q^{\infty}+ 2 \left( \hbar \omega_{LO} f^2_q- 2 M_q
f_q \right). \label{9r}
\end{equation}
Due to the electron-phonon interaction, the effective potential
gets reduced in comparison with the bare repulsive one. Moreover,
in the strong electron-phonon coupling regime ($\alpha>6$, not
analyzed in this paper), $V_q^{eff}$ can also present a negative
sign for some values of the wave-vector $q$ (characteristic of an
attractive interaction), thus  favoring the stability of
bipolaronic \cite{bipo1} or charge-ordered phases \cite{giulio}
different from the normal state considered in this paper.
Therefore, the unitary transformation $U$ given in Eq. (\ref{5r})
takes into account very relevant effects due to the
electron-phonon coupling, that is both the band shift in Eq.
(\ref{8r}) and the effective polaron-polaron potential in Eq.
(\ref{9r}), which in fact depend on the phonon distribution
function $f_q$. Other renormalization effects can be determined
analyzing the role played by further interaction terms of the
transformed Hamiltonian $\tilde{H}$.

In Eq.  (\ref{6r}), the residual polaron-phonon interaction term $H_{pol-ph} $ consists of two
contributions:
\begin{equation}
H_{pol-ph} =H_{pol-ph}^{(1)} + H_{pol-ph}^{(2)},
\label{10r}
\end{equation}
where $H_{pol-ph}^{(1)} $ has the same form as  Eq. (\ref{2r}), with $M_q$ replaced by the
following effective matrix element $M_q^{eff}$
\begin{equation}
M_q^{eff}=M_q - \hbar \omega_{LO} f_q,
\label{11r}
\end{equation}
therefore, as expected, the resulting polaron-phonon vertex is
reduced in comparison with the bare electron-phonon one. In Eq.
(\ref{10r}), $H_{pol-ph}^{(2)}$ is more complex being
\begin{equation}
H_{pol-ph}^{(2)} =\frac{1}{\sqrt{V}}
\sum_{{\bf k},{\bf q}, \sigma} N_{{\bf k},{\bf k}+{\bf q}} c^{\dagger}_{{\bf k}+{\bf q},\sigma} c_{{\bf k},\sigma} \left( a_{{\bf q}}-a^{\dagger}_{-{\bf q}}  \right),
\label{12r}
\end{equation}
where the electron-phonon matrix element $N_{{\bf k},{\bf k}+{\bf q}}$ is
\begin{equation}
N_{{\bf k},{\bf k}+{\bf q}} =   \frac{\hbar^2 f_q}{2m}  {\bf q} \cdot ({\bf q}+2 {\bf  k})=  \frac{\hbar^2 f_q }{2m} ( {\bf  k} +{\bf q}- {\bf  k} ) \cdot
( {\bf  k} +{\bf q}+{\bf  k}).
\label{12rbis}
\end{equation}
Actually,  $H_{pol-ph}^{(2)}$ derives from the unitary
transformation of the kinetic energy in Eq. (\ref{0r1}), and it is
not a function of the position operator but of the momentum
operator of phonons. Moreover, the polaron-phonon vertex in
$H_{pol-ph}^{(2)}$ does not simply depend on the phonon momentum
${\bf q}$, but it is a function of the incoming vector ${\bf k}$
and the outgoing vector ${\bf k}+ {\bf q}$. In the following
sections, we will find that the effects due to the term
$H_{pol-ph}^{(2)}$ on the spectral properties of large polarons in
the intermediate coupling regime are not negligible in comparison
with those due to the term $H_{pol-ph}^{(1)}$ when the charge
density is low.

Finally, in Eq. (\ref{6r}), the polaron-two phonon interaction
term $H_{pol-2ph} $ describes the interaction between phonons
mediated by polarons. Like $H_{pol-ph}^{(2)}$, the Hamiltonian
$H_{pol-2ph}$ derives from the unitary transformation of the
kinetic energy in Eq. (\ref{0r}).

In order to pursue the theoretical approach, one has to evaluate
the variational function $f_q$ which determines the parameters of
$\tilde{H}$. At zero temperature, $f_q$ is calculated through a
variational scheme minimizing the ground state energy $E_0$ of the
system with $N$ charge carriers. In the regime of weak to
intermediate electron-phonon coupling, the ground state
wave-function $|\Psi_0 \rangle$ of the original Hamiltonian $H$ in
Eq. (\ref{0r}) is given in terms of the unitary transformation $U$
in Eq. (\ref{5r}) in the following way:
\begin{equation}
|\Psi_0 \rangle= U |\Psi_{pol} \rangle | 0 \rangle_{ph},
\label{16r}
\end{equation}
where $|\Psi_{pol} \rangle$ is the ground state of the
many-polaron Hamiltonian (\ref{18r}) and $ | 0 \rangle_{ph}$ is
the phonon vacuum. The minimization of the ground state energy
$E_0$ provides the following form of the function $f_q$
\cite{Lemmens}:
\begin{equation}
f_q=\frac{M_q}{\hbar \omega_{LO} + \frac{\hbar^2 q^2}{2 m S^{eff}_q}},
\label{19r}
\end{equation}
where $S^{eff}_q$ is the static structure factor related to
$|\Psi_{pol} \rangle$. We notice that there is a recoil term in
Eq. (\ref{19r}), which has the same form as the Bijl-Feynman
expression for the excitations in liquid helium IV \cite{Lemmens}.
In the case of a single polaron, one gets the limit $S^{eff}_q
\rightarrow 1$.

Within the Hartree-Fock approximation for $H_{pol}$ \cite
{Lemmens},  $|\Psi_{pol} \rangle=D$, where $D$ is the Slater
determinant of $N$ free polarons. Therefore, the static structure
factor only corresponds to $S^{(0)}_q$, that of $N$ free fermions:
$S^{eff}_q=S^{(0)}_q \simeq q$ for small $q$. However, the
structure factor must increase more slowly for small $q$
\cite{Giuliani}. Indeed, the severe suppression of long wavelength
density fluctuations due to the long-range interaction is
completely neglected within the Hartree-Fock approximation.

In order to include the effects of $H_{pol-pol}$ in Eq.
(\ref{18r}) beyond the Hartree-Fock approximation considered in
the literature \cite{Lemmens}, in this paper, we have used the
approach based on the variational wave-function proposed by
Gaskell \cite{Gaskell} for the treatment of charge correlations at
the level of $RPA$. Indeed, this approach  is based on a
Slater-Jastrow wave-function, therefore the wave-function
$|\Psi_{pol} \rangle$ for the many-polaron Hamiltonian in Eq.
(\ref{18r}) is expressed as
\begin{equation}
|\Psi_{pol} \rangle=  \exp \left[ - \sum_{{\bf q}} u_q \rho_{\bf
q} \rho^{\dagger}_{\bf q}  \right] D , \label{20r}
\end{equation}
where the exponential Jastrow term, acting on the Slater
determinant $D$, depends on an additional variational function
$u_q$ which controls charge fluctuations. Numerical approaches,
such as Monte Carlo methods, have shown that the Gaskell
wave-function takes into account almost completely the
two-particle correlations providing a very accurate description of
fermonic charge liquids \cite{Ceperley,Martin}.

The Gaskell approach provides the following Jastrow function $u_q$
\begin{equation}
u_q= \frac{1}{4} \left( \frac{1}{S^{eff}_q}- \frac{1}{S^{(0)}_q}
\right),
\end{equation}
which is expressed in terms of the interacting polaron structure
factor $S^{eff}_q$ related to the free polaron structure factor
$S^{(0)}_q$ by the following equation \cite{Gaskell}:
\begin{equation}
\frac{1}{S^{eff}_q}= \sqrt{ \frac{1}{\left[S^{(0)}_q\right]^2}+
\frac{4  m n }{\hbar^2 q^2} V_q^{eff}}. \label{23r}
\end{equation}
Since $S_q^{eff}$ depends on the polaron-polaron potential
${V^{eff}_q}$ which, in turn, is a function of the distribution
function $f_q$, the two equations (\ref{19r}) and (\ref{23r}) have
to be self-consistently solved not only as a function of the
particle density, but also of the strength of the electron-phonon
coupling. Indeed, this approach takes into account on the same
footing the polaron formation and the screening due to a finite
density of charge carriers. Moreover, in the limit $q \ll k_F$,
the structure factor recovers the Bijl-Feynman formula
\cite{Giuliani} indicating that, within the RPA-Gaskell approach,
the behavior for small $q$ is beyond the poor results given by the
Hartree-Fock approximation. Actually, when polaronic effects are
neglected ($f_q=0$), Eq. (\ref{23r}) provides a structure factor
where electron-electron interactions are exactly treated in the
limit of small $q$.

In this paper, we assume the polaron radius $R_p$ in Eq.
(\ref{4r}) as unit length, therefore all the wave-vectors will be
expressed in terms of the inverse of $R_P$. Furthermore, the
energy $\hbar \omega_{LO}$ ($\simeq 100$ meV in STO based-systems)
is assumed as energy unit. Since large ranges of particle density
$n$ will be considered in this paper, we assume $10^{22}$
$cm^{-3}$ as a reference: $n=n_{22} \cdot 10^{22}$ $cm^{-3}$, that
is the particle density will be expressed in terms of $n_{22}$.
Therefore, one can easily find the order of magnitude for the
Fermi wave-vector $k_F$ and energy $E_F$:
\begin{equation}
k_F R_P=2.79 \cdot n_{22}^{\frac{1}{3}}, \ \ \frac{E_F}{\hbar
\omega_{LO}}=8.46 \cdot n_{22}^{\frac{2}{3}}   \label{23r_bis}
\end{equation}
For the density $n=10^{19}$ $cm^{-3}$, $k_F \simeq 0.067$
$\AA^{-1}$, and $E_F \simeq 8.46$ $meV$, values smaller than those
typical of simple metals. In the following, we will show that a
proper treatment of particle-particle correlations is able to
provide a description of incipient screening effects in this
regime of rather low particle density. Actually, STO-based systems
present a lot of interesting properties, such as
superconductivity, for so low densities that $E_F < \hbar
\omega_{LO}$, the so-called anti-adiabatic regime
\cite{levy,Marel}. In this paper, the analysis will focus on this
regime of particle densities where the variational approach is
able to provide a very accurate description for static quantities
of the normal state.

It is possible to relate the bulk three dimensional density to the
electron density  of two-dimensional gases at the STO surface or
at the LAO/STO interface. In fact, following Ref.
\cite{dubroka,Salluzzo}, the two-dimensional density $n_{2D}$ of
LAO/STO samples can be obtained by the volume carrier density $n$
by considering the effective thickness $d$ at the interface to be
less than $10$ nm. In this paper, we assume $d=6$ nm: $n_{2D}=n
\cdot d= n_{22} \cdot 6 \cdot 10^{15}$ $cm^{-2}$. Therefore, for
the volume density $n=2 \cdot 10^{19}$ $cm^{-3}$, $n_{2D}$ is of
the order of $1.2 \cdot 10^{13}$ $cm^{-2}$, which is the reference
density in quasi two-dimensional STO-based systems.

%In the following sections, the density variations in terms of
%$n_{22}$ will be relevant for the analysis of both three- and
%two-dimensional STO-based systems.

\subsection{Results of the variational approach}

%%%%%%%%%%%%%%%%%%%%%%%%%%%%%%
\begin{figure}[t]
\begin{center}
\includegraphics[angle=-90,scale=0.29]{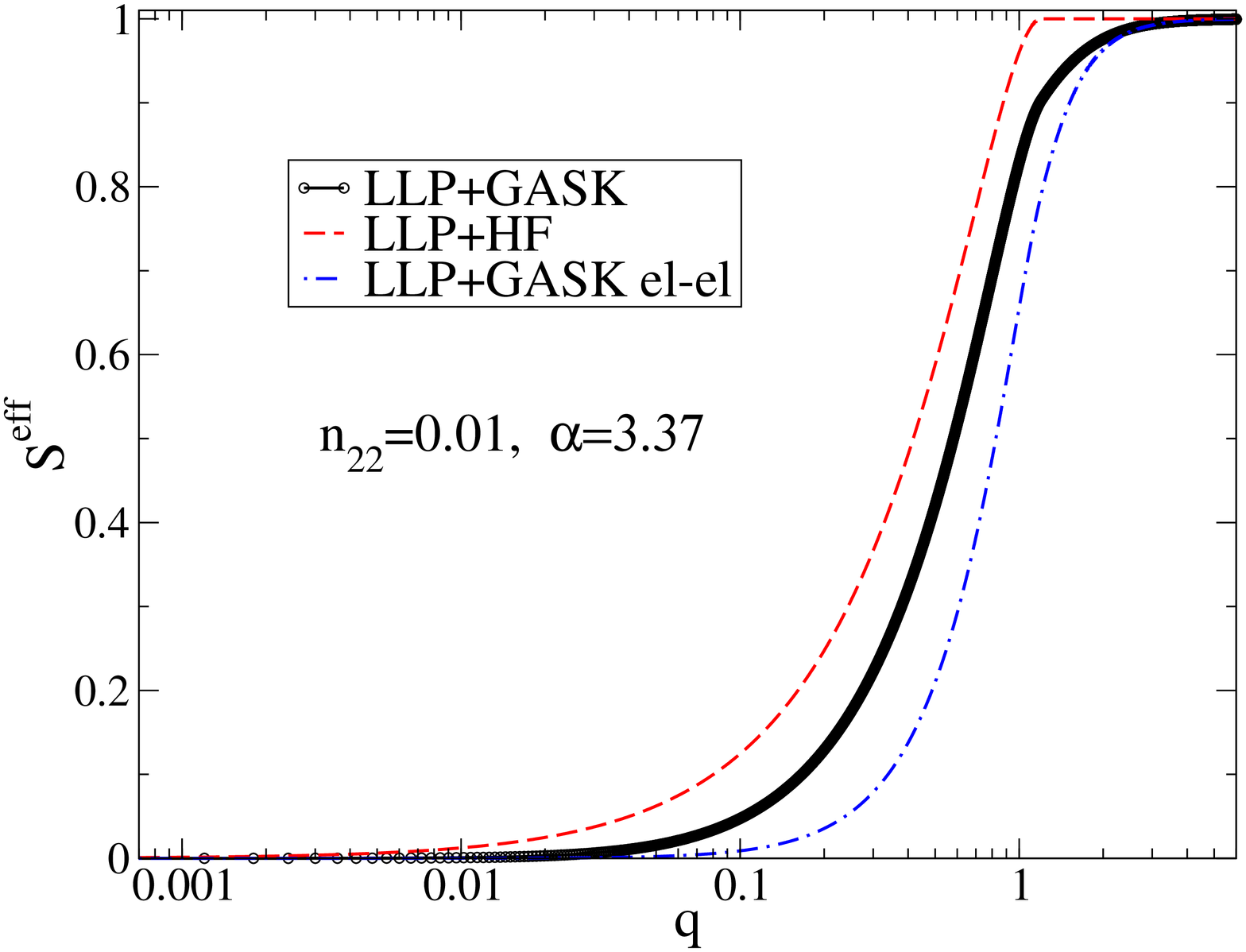} \\
\includegraphics[angle=-90,scale=0.29]{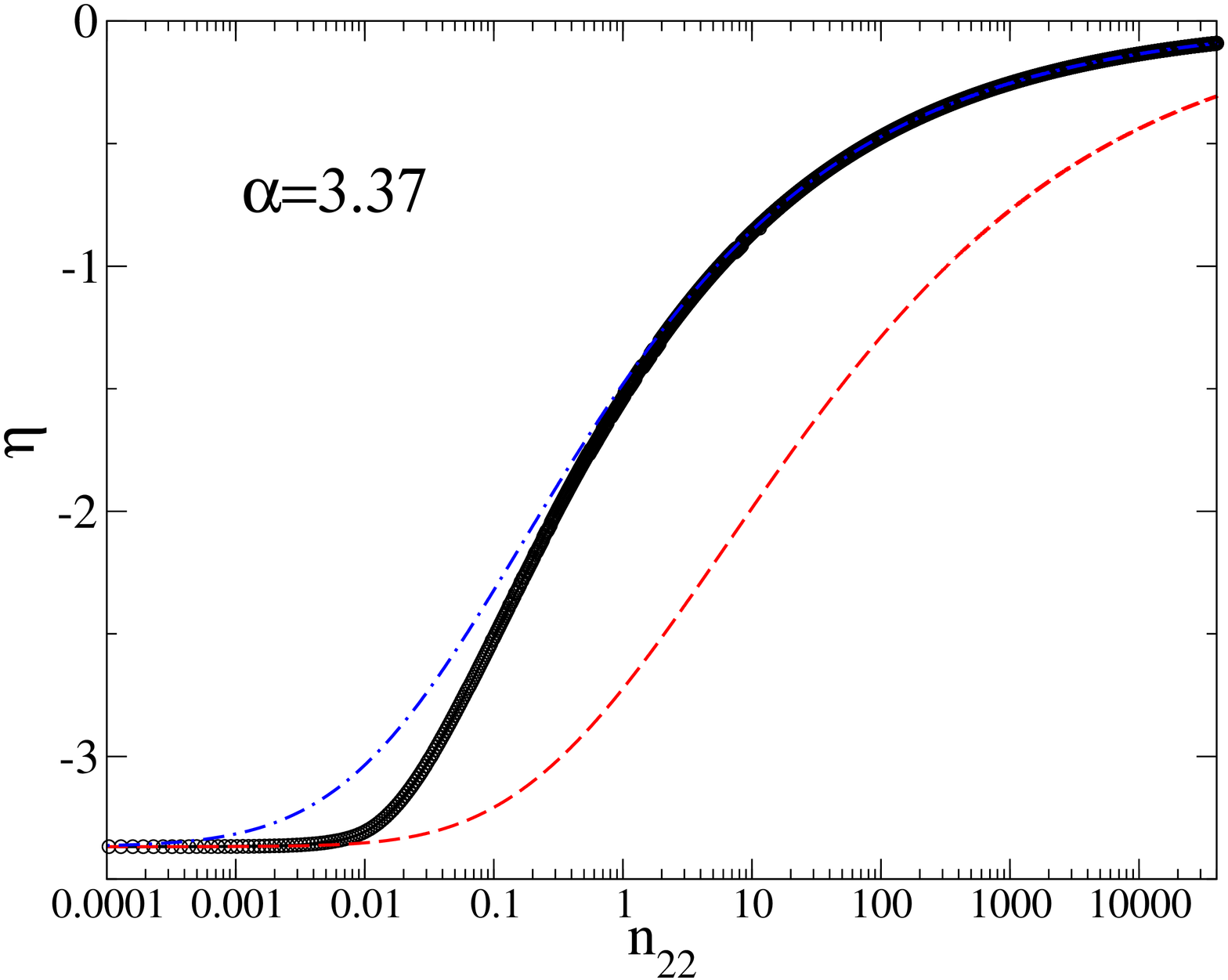}
\caption{Upper Panel: The effective structure factor $S_q^{eff}$
as a function of the wave-vector $q$ (in units of $1/R_P$) at
$n_{22}=0.01$. Lower Panel: the polaronic band shift $\eta$ (in
units of $\hbar \omega_{LO}$) as a function of $n_{22}$. In both
panels, electron-phonon coupling constant $\alpha=3.37$ and
different approaches: $LLP+GASK$ (black solid line) stands for
many-body LLP approach with Gaskell treatment of polaron-polaron
interactions, $LLP+GASK$ $el-el$ (blue dash-dot line) for the same
neglecting polaronic correlations, $LLP+HF$ (red dash line) for
many-body LLP approach with Hartree-Fock treatment of
polaron-polaron interactions.} \label{fig1}
\end{center}
\end{figure}
%%%%%%%%%%%%%%%%%%%%%%%%%%%%%}

In this subsection, we will analyze the behavior of many static
quantities calculated by means of the variational approach.

We start from the effective structure factor $S_q^{eff}$, which
represents one of the relevant quantities for the variational
approach proposed in this paper. In the upper panel of Fig.
\ref{fig1}, we plot $S_q^{eff}$ as a function of the wave-vector
$q$ comparing different approaches for the treatment of
polaron-polaron interactions. In the case of Hartree-Fock approach
($LLP+HF$ curve in the upper panel of Fig. \ref{fig1}), the
structure factor increases quite fast for small $q$, and it
presents a discontinuity at $q= 2 k_F$. On the other hand, the
structure factor obtained within the Gaskell approach ($LLP+GASK$
curve in the upper panel of Fig. \ref{fig1}) increases more slowly
for small $q$ indicating that charge correlations are accurately
treated for small values of $q$. Actually, in the limit $q \ll
k_F$, the structure factor recovers the Bijl-Feynman formula
\cite{Giuliani}
\begin{equation}
S^{eff}_q \simeq \frac{\hbar q^2}{2 m
\Omega_{PP}(q)}=\frac{\hbar}{n V_q^{eff}} \frac{\omega_{PP}^2}{2
\Omega_{PP}}, \label{25rr}
\end{equation}
where $\Omega_{PP}(q)$ is the frequency of the polaronic plasmon
\cite{alex1,rubano} given by
\begin{equation}
\Omega_{PP}(q) \simeq \omega_{PP} \left( 1+ \frac{2}{9}
\frac{v_F^2 q^2}{\omega_{PP}^2}   \right) , \label{25rrnnn}
\end{equation}
with $v_F=\hbar k_F /m$ the Fermi velocity. As discussed below,
$V_q^{eff}$ is always proportional to $1/q^2$ for small $q$,
therefore, in the limit $q \rightarrow 0$, the polaronic plasmon
tends to the constant value $\omega_{PP}$ given by
\begin{equation}
\omega_{PP}= \lim_{q \rightarrow 0} \sqrt{  \frac{n q^2}{m}
V_q^{eff}}. \label{24rr}
\end{equation}
Indeed, in the limit of small $q$, the effective structure factor
remains quadratic as a function of $q$ confirming that the normal
state of many  interacting large polarons is a charged Fermi
liquid \cite{Giuliani}. In the absence of polaronic effects
(distribution phonon function $f_q=0$),  $\omega_{PP}$ concides
with the plasmon frequency $\omega_P^{\infty}$ (the square of the
electron charge is screened by $\epsilon_{\infty}$), whose order
of magnitude is given by
\begin{equation}
\frac{\omega_P^{\infty}}{ \omega_{LO}} \simeq 11.6 \left( n_{22}
\right)^{\frac{1}{2}}. \label{plasminf}
\end{equation}
Therefore, even for the low particle density $n=10^{20}$
$cm^{-3}$, $\omega_P^{\infty}$ is larger than $\omega_{LO}$.
Actually, as discussed in the next section, the main contributions
to the spectral properties will come from the fermion scattering
with optical phonons. Finally, as reported in Eq. (\ref{25rrnnn}),
for $f_q=0$, the plasmon dispersion, quadratic as a function of
the wave-vector $q$, is almost coincident with that obtained
within RPA approach \cite{Giuliani,mahan}.

%%%%%%%%%%%%%%%%%%%%%%%%%%%%%%
\begin{figure}[t]
\begin{center}
\includegraphics[angle=-90,scale=0.29]{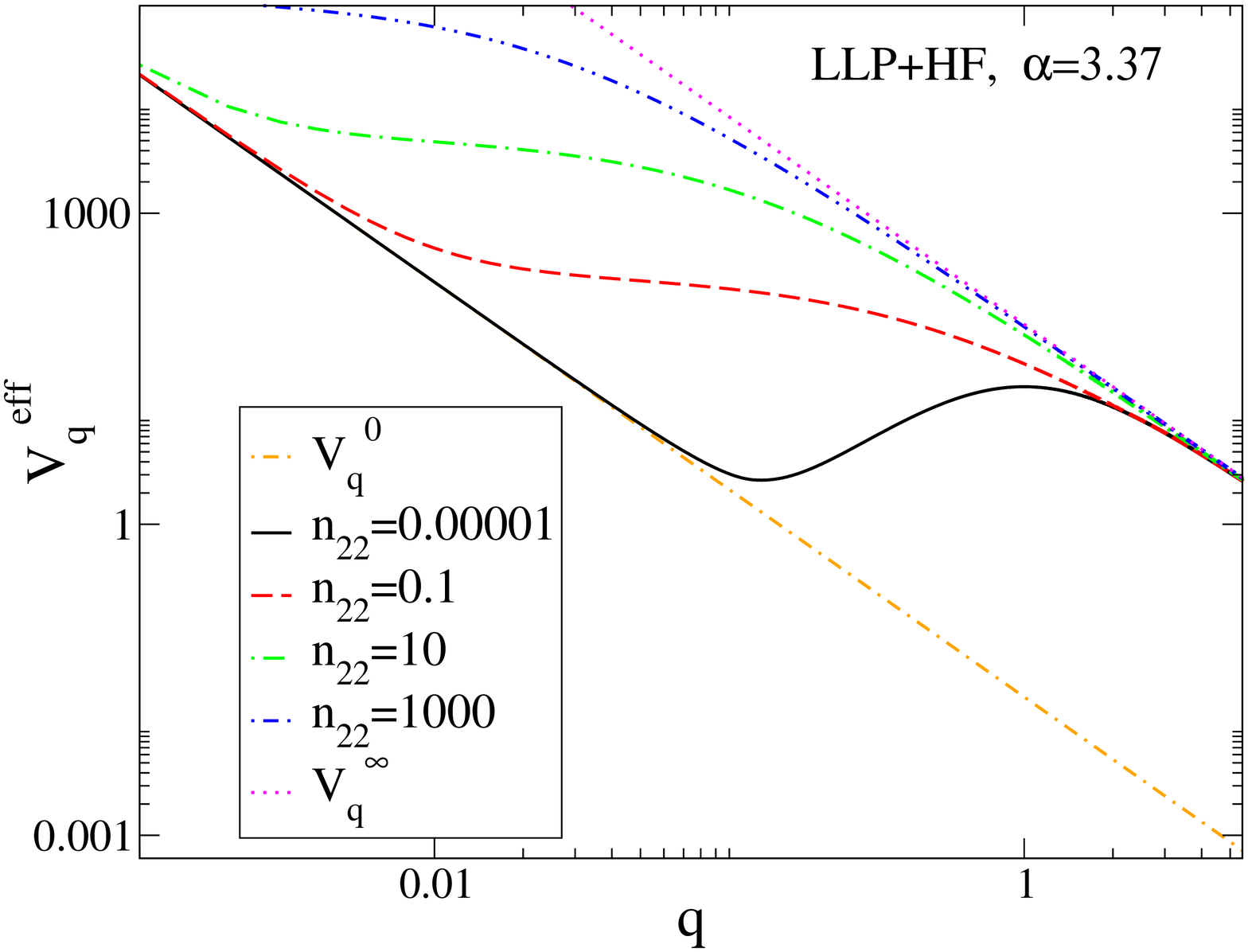} \\
\includegraphics[angle=-90,scale=0.29]{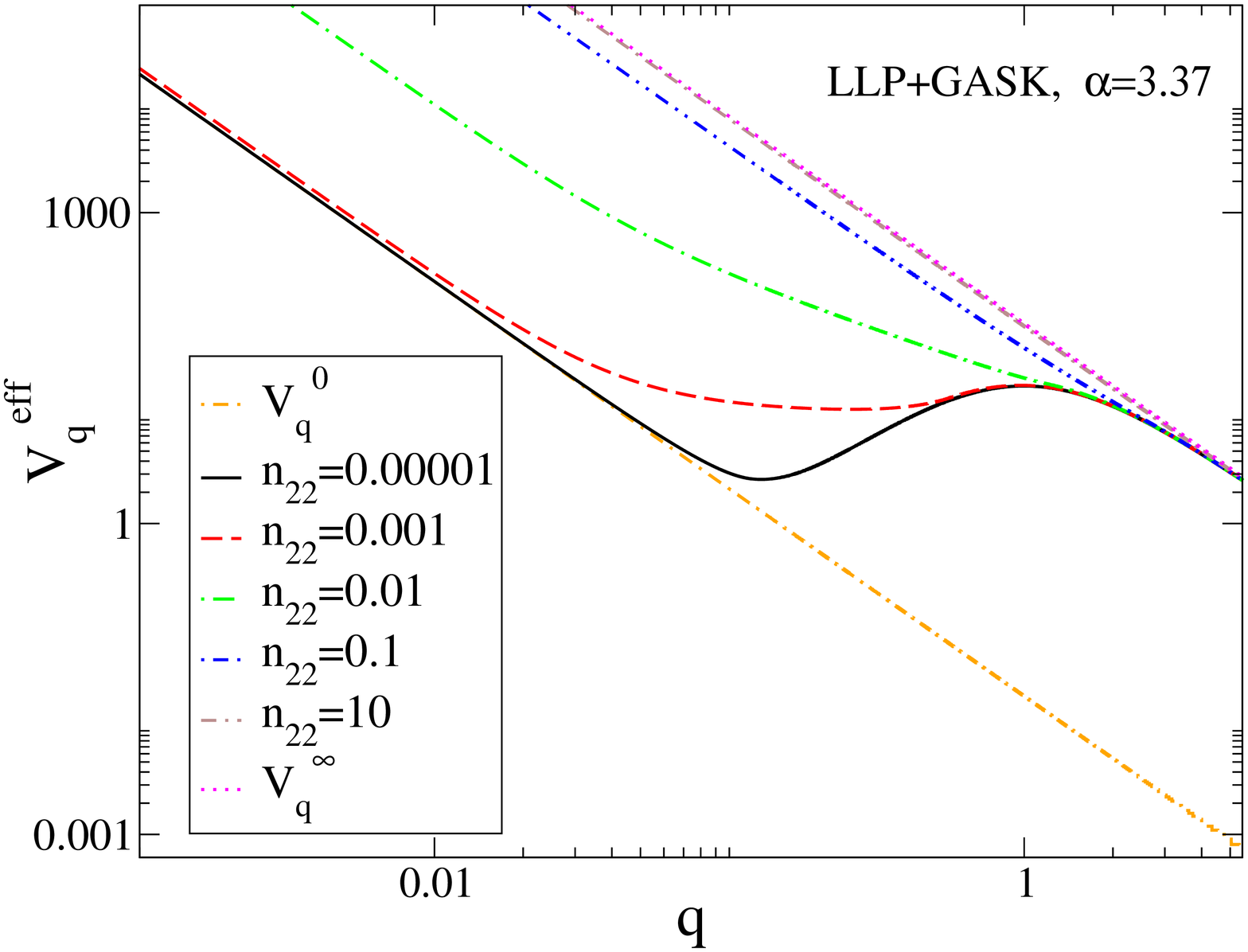}
\caption{The effective polaron-polaron potential $V_q^{eff}$ (in
units of $\hbar \omega_{LO} R_P^3$) as a function of the
wave-vector $q$ (in units of $1/R_P$) for different particle
densities at $\alpha=3.37$. $V_q^0$ is the bare Coulomb potential
screened by the static dielectric constant $\epsilon_0$, while
$V_q^{\infty}$ is the bare Coulomb potential screened by the high
frequency dielectric constant $\epsilon_{\infty}$. Upper Panel:
$LLP+HF$ stands for many-body LLP approach with Hartree-Fock
treatment of polaron-polaron interactions. Lower Panel: $LLP+GASK$
stands for many-body LLP approach with Gaskell-RPA treatment of
polaron-polaron interactions.} \label{fig2}
\end{center}
\end{figure}
%%%%%%%%%%%%%%%%%%%%%%%%%%%%%

In order to emphasize the effects of the electron-phonon coupling
on the stucture factor, in the upper panel of Fig. \ref{fig1}, we
report  $S_q^{eff}$ when only electron-electron interactions are
taken into account within the Gaskell approach neglecting
polaronic formation ($LLP+GASK$ $el-el$ in figure corresponding to
the distribution phonon function $f_q=0$ in Eq. (\ref{23r})). We
remark that the self-consistent solution of the structure factor
is necessary in the presence of polaronic effects for low
densities. Indeed, for $n_{22}=0.01$ (shown in the upper panel of
Fig. \ref{fig1}), the structure factor including polaronic
correlations shows a behavior intermediate between the
Hartree-Fock case and that obtained  at $f_q=0$.

For many interacting polarons, the phonon function distribution
$f_q$ given in Eq. (\ref{19r}) shows a behavior strongly dependent
on the properties of $S^{eff}_q$. Within the Gaskell approach, for
small values of $q$, using Eq. (\ref{19r}) and Eq. (\ref{25rr}),
one gets
\begin{equation}
f_q \simeq \frac{M_q}{\hbar \omega_{LO}+ \hbar \omega_{PP}}.
\label{25rrr}
\end{equation}
Then, for low particle densities such that $\omega_{PP} \ll
\omega_{LO}$, in the limit of small $q$, $f_q\simeq M_q/\hbar
\omega_{LO}$, the characteristic distribution function of the
single polaron. Otherwise, for high particle densities such that
$\omega_{PP} \gg \omega_{LO}$, in the limit of small $q$,
$f_q\simeq M_q /\hbar \omega_{PP} $. Therefore, polaronic effects,
quantified by $f_q$, progressively decrease  with increasing
particle density. Finally, for large $q$, $f_q$ goes as $1/q^3$.

The distribution function $f_q$ determines the polaron shift
$\eta$ defined in Eq. (\ref{8r}). In the lower panel of Fig.
\ref{fig1}, we plot $\eta$ as a function of the particle density
by using different treatments of the polaron-polaron interactions.
In particular, in the limit of small densities, as expected,
$\eta=-\alpha \hbar \omega_{LO}$ \cite{mahan}. With increasing the
density, $\eta$ becomes less negative, and, in the limit of high
densities, it goes to zero. As shown in the lower panel of Fig.
\ref{fig1}, the Hartree-Fock approach completely fails to describe
the behavior of $\eta$ providing a systematic overestimation of
the modulus of $\eta$. Actually, polaron-polaron correlations has
to be necessarily included in order to correctly describe the
renormalization of the polaronic band, which, as discussed in
Appendix B, will provide the correction to the chemical potential
$\mu$ due to many-body interactions. Finally, we point out that
polaronic effects are relevant for densities up to $n_{22}=1$,
which, as discussed in the following sections, can be considered
as the cut-off density for the manifestation of electron-phonon
effects.

Another interesting outcome of our approach is the behavior of the
polaron-polaron potential $V_q^{eff}$, defined in Eq. (\ref{9r}),
comparing Hartree-Fock approximation and Gaskell approach. This
comparison confirms that the Hartree-Fock approach does not
correctly describe the behavior of static quantities with
increasing particle density.

In the upper panel of Fig. \ref{fig2}, we plot $V_q^{eff}$ when
the Hartree-Fock approximation is used to determine the phonon
distribution function $f_q$. Actually, for small values of $q$, in
the limit of low density, $V_q^{eff}$ goes toward
$V_q^0=V_q/\epsilon_0$, that is the Coulomb potential screened by
the static dieletric function $\epsilon_0$ \cite{bipo1}. In the
limit of large values of $q$, $V_q^{eff}$ tends to $V_q^{\infty}$,
the bare Coulomb potential. With increasing the density, one
expects a crossover towards a regime where $V_q^{eff}$ tends to be
more similar to $V_q^{\infty}$ for smaller values of $q$. However,
even for the high density $n_{22}=1000$, this crossover is not
complete within the Hartree-Fock approach.  Therefore, in the next
section, we will analyze dynamic quantities including always
correlations beyond Hartree-Fock approximation.

In the lower panel of Fig. \ref{fig2}, we report $V_q^{eff}$ when
the Gaskell approach is used to determine the phonon distribution
function $f_q$. In particular, in the limit of large $q$,  by
using Eq. (\ref{2rbis}), at the leading order, $V_q^{eff} \simeq
V_q^{\infty}$. Instead, for small values of the wave-vector $q$,
by using Eq. (\ref{25rr}), one gets
\begin{equation}
V_q^{eff} \simeq  V_q^{\infty} +\left( V_q^0-V_q^{\infty} \right)
\left[ 1- \frac{\omega_{PP}^2}{(\omega_{LO}+\omega_{PP})^2}
\right]. \label{25rrrr}
\end{equation}
For low particle densities such that $\omega_{PP} \ll
\omega_{LO}$, in the limit of small $q$, $V_q^{eff}\simeq V_q^0$.
Otherwise, for large particle densities such that $\omega_{PP} \gg
\omega_{LO}$, in the limit of small $q$, $V_q^{eff}\simeq
V_q^{\infty}$. Actually, neglecting polaronic effects ($f_q=0$),
from Eq. (\ref{plasminf}), one can determine the crossover density
$n_{22}^c$ defined such that $\omega_{P}^{\infty} \simeq
\omega_{LO}$: $n_{22}^c\simeq 0.01$. In fact, as shown in the
lower panel of Fig. \ref{fig2}, at $n_{22}=0.01$, $V_q^{eff}$
seems to be intermediate between $V_q^{\infty}$ and $V_q^0$ for
low values of $q$.

In contrast with Hartree-Fock approximation, we notice that,
within the Gaskell approach, the crossover of $V_q^{eff}$ from
$V_q^0$ to $V_q^{\infty}$ is very rapid with increasing the
particle density. Indeed, as shown in the lower panel of Fig.
\ref{fig2}, already at $n_{22}=0.1$, the crossover is almost
complete. Really, at $n_{22}=10$, $V_q^{eff}$ is practically
identical to $V_q^{\infty}$. Apparently, as discussed in the next
section, in order to calculate the spectral properties, one needs
to include a proper screening of the polaron-polaron interactions
through the effects of excited states.

\section{Spectral properties}
In the previous section,  we have characterized relevant terms of
the transformed Hamiltonian $\tilde{H}$ in Eq. (\ref{6r}). In this
section, the approach used to calculate the spectral properties
will be based upon an accurate many-body perturbation theory of
the interaction terms in the transformed Hamiltonian $\tilde{H}$.
The small effects due to the anharmonic term $H_{pol-2ph}$ in Eq.
(\ref{6r}) are neglected focusing on the effects of the terms
$H_{pol-pol}$ in Eq. (\ref{18r}) and $H_{pol-ph}$ in Eq.
(\ref{10r}). The electron spectral properties of the system are
calculated at zero temperature for a finite density $n$ of charge
carriers in the regime from weak to intermediate electron-phonon
coupling constant. In the absence of polaronic effects
(distribution phonon function $f_q=0$), this theory recovers the
approach valid in the perturbative regime of electron-electron and
electron-phonon coupling \cite{mahan}.

After performing the  canonical transformation given in Eq.
(\ref{5r}), within the perturbative approach, the two-point
electron correlation function can be disentangled into polaronic
and phononic contributions
\cite{alex1,manga1,manga2,charge1,charge2} yielding the following
electronic Green's function ${\mathcal G} \left( {\bf k},i \hbar
k_n \right)$ in fermionic Matsubara frequencies $k_n$

\begin{eqnarray}
&& { \mathcal G} \left( {\bf k},i \hbar k_n \right)=
 e^{-S} { \mathcal G}_{pol} \left( {\bf k},i \hbar k_n \right) +
\nonumber \\
&& \frac{e^{ -S } }{V} \sum_{\bf{k}_1} \sum_{l=1}^{\infty}
\frac{1}{l!} F_l({\bf k}-{\bf k}_1) [ {\mathcal G}_{pol} \left(
k_1,i \hbar k_n+l \hbar \omega_{LO}  \right) n_F( \xi_{k _1})
 \nonumber \\
&& +  {\mathcal G}_{pol} \left( k_1,i \hbar k_n-l \hbar
\omega_{LO} \right) \{1- n_F( \xi_{k_1}) \} ], \label{136r}
\end{eqnarray}
where $S=K({\bf r}=0)$, with the function $K({\bf r})$ given in
terms of the distribution function $f_q$ as
\begin{equation}
K({\bf r})= \frac{1}{V} \sum_{{\bf q}} f^2_q \cos{ \left( {\bf q}
\cdot {\bf r} \right) },
\label{137r}
\end{equation}
the function $F_l({\bf p})$ is defined as
\begin{equation}
F_l({\bf p})= \int d {\bf r} e^{-i  {\bf p} \cdot {\bf r} } \left[
K({\bf r}) \right]^l,
\label{138r}
\end{equation}
$\xi_k=\epsilon_k-\mu$, with $\epsilon_k$ polaronic band given in
Eq. (\ref{7r}), and $n_F(E)=\theta(\mu-E)$ is the Fermi
distribution function at zero temperature, with $\mu$ chemical
potential. In Eq. (\ref{136r}), ${ \mathcal G}_{pol} \left( {\bf
k},i \hbar k_n \right)$ is the polaron Green's function, defined
starting from the transformed Hamiltonian  $\tilde{H}$ of Eq.
(\ref{6r}).

We point out that two physically distinct terms appear in Eq.
(\ref{136r}): the coherent and the incoherent one
\cite{alex,ranni}. The first term derives from the coherent motion
of electrons and their surrounding phonon cloud. Without many-body
corrections in the Hamiltonian  $\tilde{H}$ of Eq. (\ref{6r}), the
first term of the spectral function derived from Eq. (\ref{136r})
represents the purely polaronic band contribution and shows a
delta behavior. This coherent term is controlled by the
exponential $e^{-S}$, which will represent the most important
contribution to the spectral weight of the Green's function. On
the other hand, the second term in Eq. (\ref{136r}) describes the
possibility of changing the number of phonons in the phonon cloud
during the electron motion. This is confirmed by the presence of
phonon replicas and the sum over all momenta. This second term
provides the incoherent contribution and spread over a wide energy
range.

The polaronic Green's function ${ \mathcal G}_{pol} \left( {\bf
k},i \hbar k_n \right)$ in Eq. (\ref{136r}) is related to the
polaronic self-energy $ \Sigma_{pol} \left( {\bf k},i \hbar k_n
\right)$ by means of the Dyson equation \cite{mahan}:
\begin{equation}
{ \mathcal G}_{pol} \left( {\bf k},i k_n \right)= \frac{ {
\mathcal G}^{(0)}_{pol} \left( {\bf k},i k_n \right)  } { 1- {
\mathcal G}^{(0)}_{pol} \left( {\bf k},i k_n \right) \Sigma_{pol}
\left( {\bf k},i \hbar k_n \right)},
 \label{31rzz}
\end{equation}
where ${ \mathcal G}^{(0)}_{pol} \left( {\bf k},i k_n \right)$ is
the free polaron Green's function. Indeed, the introduction of the
polaronic self-energy allows to include directly additional
dampings and energy renomalizations for the large polarons
improving the approximations for the calculation of the spectral
properties \cite{alex1,manga1,manga2,charge1,charge2}. We have
checked that the self-energy does not change the spectral
properties in a considerable manner, although it allows to
eliminate the delta behavior in the expression of the coherent
term of the spectral function.

In order to determine the polaronic self-energy, we have to
evaluate the total dynamic polaron potential ${\cal
W}^{(tot)}_{{\bf k}, {\bf p}, {\bf q}} (i \hbar q_n)$ in bosonic
Matsubara frequencies $q_n$, which, due to the complex
polaron-phonon vertex of $\tilde{H}$, depends not only on the
phononic momentum $\hbar {\bf q}$, but also on both the incoming
polaronic momenta $\hbar {\bf k}$ and $\hbar {\bf p}$. Actually,
with increasing the particle density, it is necessary to properly
screen the polaron-polaron interaction and the polaron-phonon
couplings in the limit of small $q$. In this paper, the screening
is introduced by  the dielectric function $\epsilon_q (i \hbar
q_n)$ in bosonic Matsubara frequencies $q_n$, which, in general,
includes contributions from electron-electron and electron-phonon
interactions \cite{mahan}. Therefore, the total potential ${\cal
W}^{tot}_{{\bf k}, {\bf p}, {\bf q}} (i \hbar q_n)$ is derived as
\begin{equation}
{\cal W}^{tot}_{ {\bf k}, {\bf p}, {\bf q} } (i \hbar q_n)= \frac{
{\cal W}^{(0)}_{{\bf k}, {\bf p}, {\bf q}} (i \hbar q_n)}
{\epsilon_q (i \hbar q_n) }, \label{31r}
\end{equation}
where the bare potential ${\cal W}^{(0)}_{{\bf k}, {\bf p}, {\bf
q}} (i \hbar q_n)$ is
\begin{equation}
{\cal W}^{(0)}_{{\bf k}, {\bf p}, {\bf q}} (i \hbar q_n)=
V^{eff}_q + {\cal W}^{(ph)}_{{\bf k}, {\bf p}, {\bf q}} (i \hbar
q_n), \label{26rzz}
\end{equation}
with $V^{eff}_q$ the static polaron potential defined in Eq.
(\ref{9r}), and ${\cal W}^{(ph)}_{{\bf k}, {\bf p}, {\bf q}} (i
\hbar q_n)$ the phonon-mediated dynamic potential.

The phonon-mediated dynamic potential ${\cal W}^{(ph)}_{{\bf k},
{\bf p}, {\bf q}} (i \hbar q_n)$ in Eq. (\ref{26rzz}) comes from
integrating out the phonon degrees of freedom interacting with
polarons through the term $H_{pol-ph}$ of $\tilde{H}$ in Eq.
(\ref{6r}), hence, in bosonic Matsubara frequencies $q_n$,
\begin{equation}
{\cal W}^{(ph)}_{{\bf k}, {\bf p}, {\bf q}} (i \hbar q_n)=\left[
\left( M_q^{eff} \right)^2  + U^{(0)}_{{\bf k}, {\bf p}, {\bf q}}
\right] {\cal D}^{(0)}(i \hbar q_n) , \label{26r_ter}
\end{equation}
where $M_q^{eff}$ is the effective polaron-phonon matrix element
defined in Eq. (\ref{11r}), $U^{(0)}_{{\bf k}, {\bf p}, {\bf q}}$
is the term derived from $H^{(2)}_{pol-ph}$ given in Eq.
(\ref{12r})
\begin{eqnarray}
U^{(0)}_{{\bf k}, {\bf p}, {\bf q}} &=& -N_{{\bf k},{\bf k}+{\bf
q}}  N_{{\bf p},{\bf p}-{\bf q}}
\\ \nonumber &=& f_q^2\left( \frac{ \hbar^2}{2m} \right)^2 (2 {\bf k}
\cdot {\bf q} +q^2)  (2 {\bf  p} \cdot   {\bf q} -q^2),
\label{27r_ter}
\end{eqnarray}
where ${\bf k}$ and ${\bf p}$ are the incoming wave-vectors, ${\bf
k}+{\bf q}$ and ${\bf p}-{\bf q}$ the outgoing wave-vectors,
$N_{{\bf k},{\bf k}+{\bf q}} $ is the polaron-phonon matrix
element in Eq. (\ref{12rbis}), and ${\cal D}^{(0)}(i \hbar q_n)$
is the free phonon Green function in Matsubara frequencies $q_n$
\begin{equation}
{\cal D}^{(0)}(i \hbar q_n)=\frac{2 \hbar \omega_{LO}}{(i \hbar
q_n)^2- (\hbar \omega_{LO})^2}.
\label{26rbis_ter}
\end{equation}
For the calculation of the self-energy, we have to consider
\begin{equation}
U^{(0)}_{{\bf k}, {\bf k}+{\bf q}, {\bf q}} = f_q^2\left( \frac{
\hbar^2}{2m} \right)^2 \left( 2 {\bf k} \cdot {\bf q} +q^2
\right)^2, \label{27r_ter1}
\end{equation}
which is a positive quantity like $\left(M_q^{eff}\right)^2$, but
it depends on the angle between ${\bf k}$ and ${\bf q}$. We remark
that $U^{(0)}_{{\bf k}_1, {\bf k}_1+{\bf q}, {\bf q}},$, with
${\bf k}_1=k_F {\hat q}$, is equal to the coupling $T_{2,q}$,
which, in addition to $T_{1,q}=\left(M_q^{eff}\right)^2$, is
analyzed in Appendix A. As discussed in this Appendix, the two
polaron-phonon coupling terms $T_{1,q}$ and $T_{2,q}$ provide
comparable contributions to the spectral properties in the
intermediate electron-phonon coupling regime for low particle
densities.

After having evaluated the total polaron potential ${\cal
W}^{(tot)}_{{\bf k}, {\bf p}, {\bf q}} (i \hbar q_n)$, the
polaronic self-energy  to the lowest order can be obtained as
\begin{eqnarray}
\Sigma_{pol} && \left( {\bf k},i \hbar k_n \right)= \\ \nonumber
&& -\frac{1}{\beta V} \sum_{{\bf q},q_n} {\cal W}^{(tot)}_{{\bf
k}, {\bf k}+{\bf q}, {\bf q}} (i \hbar q_n)  { \mathcal
G}^{(0)}_{pol}\left( {\bf k}+{\bf q},i \hbar k_n +i \hbar q_n
\right),
\label{26r_bis}
\end{eqnarray}
where $\beta=1/k_B T$, with $k_B$ Boltzmann constant, $T$
temperature.  Making the analytic continuation $ i \hbar k_n
\rightarrow E +i \delta $, with $\delta$ infinitesimal quantity,
and the limit to zero temperature, Eqs. (\ref{136r}) and
(\ref{31rzz}) allow to evaluate the retarted electronic Green's
function $ G^{ret} ( {\bf k}, E)$ and the electronic spectral
function $A( {\bf k}, E) = -2 \Im G^{ret} ( {\bf k}, E)$, which
will be thoroughly discussed in the next subsections. We have
checked that the sum rule $ \int_{- \infty}^{+ \infty} \frac{d
E}{2 \pi} A( {\bf k}, E) =1 $ is satisfied with a tolerance of a
few per cent for all  the electron-phonon coupling regimes and
particle densities analyzed in this paper.

\subsection{Single polaron - low density regime in STO-based systems}
In the limit of very low particle density, analytic calculations
can be made to determine many contributions to the spectral
function not only within the electron-phonon perturbative regime
\cite{giulio,notaperturbazione}, but also within the LLP approach.
In particular, we focus on the spectral weight at $k=0$, indicated
with $Z_0$, which represents a relevant measure of the polaronic
character in the case of a single fermion \cite{mahan}.

\begin{figure}[t]
\begin{center}
\includegraphics[angle=0,width=8.5cm]{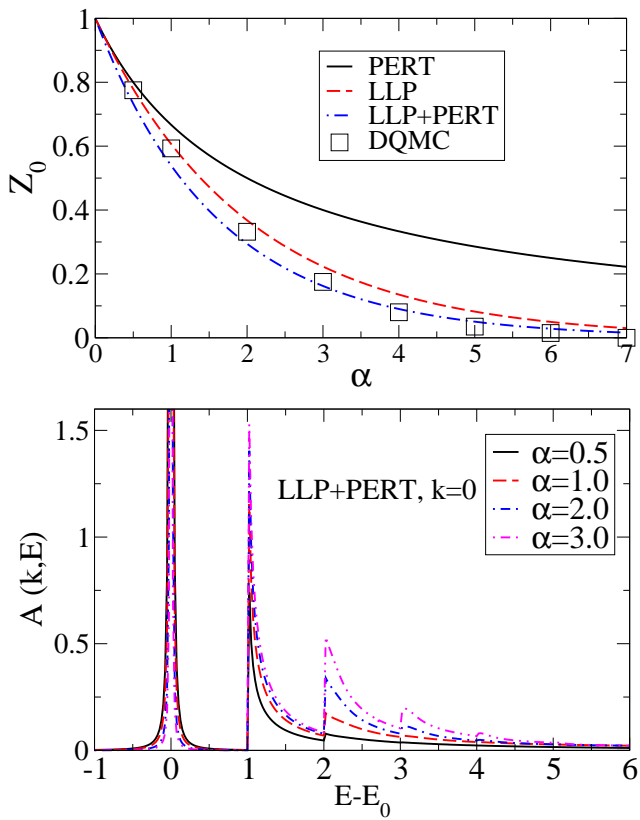}
\caption{ Upper Panel: Single fermion spectral weight at
wave-vector $k=0$ as a function of the electron-phonon coupling
constant $\alpha$ for different approaches: $PERT$ stands for
perturbative approach, $LLP$ for Lee-Low-Pines method, $LLP+PERT$
for perturbative corrections upon LLP method, $DQMC$ for
Diagrammatic Quantum Monte-Carlo data from Ref. \cite{andrei1}.
Lower Panel: Single fermion spectral function (in units of
$1/\hbar \omega_{LO})$ at wave-vector $k=0$ as a function of the
energy (related to the ground state energy $E_0$, both in units of
$\hbar \omega_{LO}$) at $k=0$ for different values of the
electron-phonon coupling constant $\alpha$ within the perturbative
method upon the LLP approach (LLP+PERT). A negligible width
$\Gamma$ has been added as an imaginary part to the polaronic
self-energy.} \label{fig4}
\end{center}
\end{figure}

The scheme  perturbative in the electron-phonon coupling provides
the following estimate for $Z_0$:
\begin{equation}
Z_0^{PERT}=\frac{1}{1+\frac{\alpha}{2}},
\label{100r}
\end{equation}
such that, for very low $\alpha$, $Z_0^{PERT}\simeq 1-\alpha/2$,
which is commonly used in the literature \cite{mahan}. In any
case, as shown in the upper panel of Fig. \ref{fig4}, the spectral
weight decreases with increasing the coupling constant $\alpha$.
In the same panel, we report the spectral weight calculated within
the LLP scheme:
\begin{equation}
Z_0^{LLP}=\exp{\left( -\frac{\alpha}{2} \right)}, \label{101r}
\end{equation}
We point out that $Z_0^{LLP}$ represents a very relevant
contribution to the exact spectral weight. Indeed, as shown in the
upper panel of Fig. \ref{fig4}, $Z_0^{LLP}$ is slightly larger
than the numerically exact results from DQMC technique.  For
$\alpha>4$, the spectral weight is smaller than $0.1$, therefore,
the spectral function derived from Eq. (\ref{136r}) is dominated
by the incoherent term. This is the reason why, in the lower panel
of Fig. \ref{fig4}, we focus on the spectral functions for
$\alpha<4$. In particular, for $\alpha=3.37$, which is the value
estimated to be relevant for STO-based systems, $Z_0$ is very
close to $0.2$, hence, the quasi-particle, the large polaron, is
still well defined. As shown in the next subsection,  this value
of $Z_0$ for $\alpha=3.37$ is in very good agreement with its
estimate from experimental data in the limit of small particle
densities \cite{devereaux}.

%%%%%%%%%%%%%%%%%%%%%%%%%%%%%%
%%%%%%%%%%%%%%%%%%%%%%%%%%%%%

In order to go beyond the LLP scheme and to determine the
polaronic self-energy $\Sigma_{pol} \left( {\bf k},i \hbar k_n
\right)$, in the case of a single fermionic particle, we only need
${\cal W}^{(ph)}$ in Eq. (\ref{26r_ter}), thus, making the limit
of very low particle density for all the quantities,  ${\cal
W}^{(tot)}= {\cal W}^{(ph)}$. Analytic calculations can be made to
determine the retarded polaronic self-energy $\Sigma_{pol}^{ret}
\left( {\bf k},E \right)$. In particular, including the lowest
order polaron-phonon corrections upon the LLP scheme, the spectral
weight at $k=0$, $Z_0^{LLP+PERT}$, is
\begin{equation}
Z_0^{LLP+PERT}=Z_0^{LLP} \cdot Z_0^{PERT}= \frac{ \exp{\left(
-\frac{\alpha}{2} \right)} }{1+\frac{\alpha}{8} }. \label{102r}
\end{equation}
Therefore, the polaron-phonon correction provides a denominator
which is similar to that obtained for the electron-phonon
perturbation theory given in Eq. (\ref{100r}). As expected, the
fraction in the denominator is smaller than that present in Eq.
(\ref{100r}) since the polaron-phonon interaction terms are
reduced in comparison with the bare electron-phonon vertex. As
shown in the upper panel of Fig. \ref{fig4}, $Z_0^{LLP+PERT}$ is
in excellent agreement with numerical DQMC data suggesting that
weak polaron-phonon corrections are effective to improve the
accuracy of the spectral properties.

Next, we analyze the polaron effective mass at $k=0$, denoted with
$m_0^{*}$, which quantifies the mass increase due to the phonon
cloud accompanying the electron  \cite{mahan}. The approach
perturbative in the electron-phonon coupling provides the
following estimate for $m_0^{*}$:
\begin{equation}
\frac{m}{m_0^{*}}= \left( 1 + \frac{\alpha}{3} \right) \cdot
Z_0^{PERT}=\frac{1 + \frac{\alpha}{3}}{1+\frac{\alpha}{2}}>
Z_0^{PERT}, \label{110r}
\end{equation}
such that, for very low $\alpha$, $\frac{m}{m_0^{*}} \simeq
1-\frac{\alpha}{6}$, which is typically used in the literature
\cite{mahan}.

In Appendix B we provide some details concerning the dispersion of
the polaron as a function of the wave-vector. In particular, as
discussed in this Appendix, the evaluation of the polaronic
self-energy in the case of a single fermion allows to estimate
also the effective mass. It is found that, for $\alpha<4$, the
effective mass at $k=0$ is well approximated by the following
expression which extends the perturbative estimate to the
intermediate coupling regime:  $\frac{m}{m_0^{*}} \simeq
1-\frac{\alpha}{6}$. Indeed, the corrections to the mass for the
Fr\"ohlich single polaron are power laws in the constant coupling
$\alpha$ within the intermediate electron-phonon coupling regime,
therefore they are weaker than those, exponential in $\alpha$,
characteristic of the spectral weight \cite{andrei1}. Moreover,
for $\alpha=3.37$, value relevant for STO-based systems,
$\frac{{m_0^{*}}}{m} \simeq 2.28$. We remark that this value is in
very good agreement with the experimental estimate
$\frac{{m_0^{*}}}{m} \simeq 2.33 $ obtained in the limit of low
particle densities \cite{devereaux}.

Finally, we focus on the spectral function at $k=0$ derived from
Eq. (\ref{136r}) in the limit of low particle density. In the
lower panel of Fig. \ref{fig4}, we plot the spectral function at
$k=0$ as a function of the energy for different values of the
electron-phonon coupling constant $\alpha$. We notice that the all
the curves show the peak-dip-hump structure characteristic of the
experimental spectral function
\cite{devereaux,cancellieri,strocov}. The peak corresponds to the
coherent term, while the hump to the incoherent contribution.
Therefore, the increase of the coupling constant $\alpha$ induces
a transfer of spectral weight towards higher energies enhancing
the number of phonon satellites. In fact, for $\alpha=0.5$,
one-phonon satellite is evident in the spectra, while, for
$\alpha=3.37$, coupling relevant for STO-based systems, at least
three phonon satellites  characterize the incoherent term. We
remark that these features of the spectrum for $\alpha=3.37$ are
in very good agreement with the electron spectral function
extracted from experiments in the limit of low particle density
\cite{devereaux}.

Summarizing, not only spectral weight and effective mass, but also
the spectral function at $k=0$ in the limit of small density are
accurately described by the approach used in this paper confirming
that the estimated value $\alpha=3.37$ is perfectly consistent
with experimental results in STO-based systems. In the next
subsections, we analyze the effects of a finite particle density
for which polaron-polaron interactions become relevant together
with polaron-phonon couplings.

\subsection{RPA-Gaskell approach - low to high density regime in STO-based systems}

We recall that the calculation of  the electronic Green's function
${ \mathcal G}\left( {\bf k},i k_n \right)$ in Eq. (\ref{136r})
requires the evaluation of the polaronic Green's function ${
\mathcal G}_{pol} \left( {\bf k},i k_n \right)$ in Eq.
(\ref{31rzz}) and its phonon replicas. In turn, the polaronic
Green's function is calculated through the polaronic self-energy
$\Sigma_{pol} \left( {\bf k},i \hbar k_n \right)$, which requires
the total dynamic polaron potential ${\cal W}^{tot}_{{\bf k}, {\bf
k}+{\bf q}, {\bf q}} (i \hbar q_n)$ in Eq. (\ref{31r}) obtained by
screening the bare polaron potential ${\cal W}^{(0)}_{{\bf k},
{\bf k}+{\bf q}, {\bf q}} (i \hbar q_n)$ given in Eq.
(\ref{26rzz}) through the dielectric function $\epsilon_q (i \hbar
q_n)$.

At the level of the Hartree-Fock approximation, in Eq.
(\ref{31r}), the dielectric function is approximated to unity,
therefore the total dynamical potential reduces to the bare one
${\cal W}^{(0)}$: ${\cal W}^{(tot)}= {\cal W}^{(0)}$. Therefore,
screening effects due to the presence of fermionic charge carriers
are not included in the dynamical potential. As expected, at the
Hartree-Fock level, the effects of a finite density $n$ of charge
carriers are not correctly taken into account. In this paper, the
calculation of the dielectric function is made extending the
Gaskell approach beyond the ground state which has been analyzed
in the previous section. Since the results are similar to the RPA
approach, we call this method RPA-Gaskell.

We have checked that a very accurate starting point to calculate
the dielectric function is the self-consistent structure factor
$S_q^{eff}$ which, for small $q$, satisfies the Bijl-Feynman
relation in Eq. (\ref{25rr}) related to the polaron plasmon
$\Omega_{PP}(q)$ in Eq. (\ref{25rrnnn}). Indeed, we recall that
the static structure factor can be obtained for $q \neq 0$ as
\begin{equation}
S_q^{eff}=\int_0^{\infty} \frac{ d \hbar \omega}{\pi}
S_q^{eff}(\hbar \omega), \label{31rn}
\end{equation}
where $S_q^{eff}(\hbar \omega)$ is the dynamic structure factor,
which is the spectral function of the retarded inverse dielectric
function $\epsilon_q^{ret}(\hbar \omega)$ \cite{mahan}:
\begin{equation}
S_q^{eff}(\hbar \omega)=- \frac{1}{n V_q^{eff}} \Im \left[
\frac{1}{\epsilon_q^{ret}(\hbar \omega)} \right] , \label{31rnn}
\end{equation}
with $\epsilon_q^{ret}(\hbar \omega)$ derived from $\epsilon_q (i
\hbar q_n)$ through the analytic continuation $q_n \rightarrow
\omega+i \delta$ ($\delta$ is an infinitesimal quantity). In order
to satisfy the Bijl-Feynman relation of Eq. (\ref{25rr}) for small
$q$, the inverse dielectric function in Matsubara frequencies must
have the following form:
\begin{eqnarray}
\frac{1}{\epsilon_q(i \hbar q_n)}&&= 1+ \frac{\hbar
\omega_{PP}^2}{2 \Omega_{PP}(q)} \times \\ \nonumber &&
\left[\frac{1}{i \hbar q_n- \hbar \Omega_{PP}(q)} -\frac{1}{i
\hbar q_n+\hbar \Omega_{PP}(q)}\right], \label{331rx}
\end{eqnarray}
with $\omega_{PP}$ the polaron plasmon at zero wave-vector given
in Eq. (\ref{24rr}). Therefore, as expected for small values of
$q$, the dielectric function is dominated by plasmon poles
\cite{hedin}, which are related to the polaron plasmon with
frequency $\Omega_{PP}(q)$. This is the dielectric function that
can be used in Eq. (\ref{31r}) to accurately screen
polaron-polaron and polaron-phonon interactions.

Since dynamic effects due to plasmons have not been identified in
the electronic spectral properties of STO-based systems, we focus
on the static dielectric function $\epsilon_q$ which, as expected,
goes as $1/q^2$ for small values of $q$:
\begin{equation}
\epsilon_q= 1+
\frac{\omega_{PP}^2}{\Omega_{PP}^2(q)-\omega_{PP}^2}=1+
\frac{q_{GFT}^2}{q^2}, \label{31rz}
\end{equation}
where $q_{GTF}$ is a generalized Thomas-Fermi wave-vector, defined
as
\begin{equation}
q_{GTF}=\frac{3}{2} \frac{\omega_{PP}}{v_F}, \label{31rza}
\end{equation}
with $v_F$ Fermi velocity. In fact, in the absence of polaronic
effects ($f_q=0$), $\epsilon_q$ is practically coincident with the
Thomas-Fermi dielectric function, which provides an accurate
screening of the long-range electron-electron potential for small
values of $q$ and is frequently used also for screening the
potential due to the electron-phonon coupling \cite{mahan}. As a
result, for the evaluation of the potential in Eq. (\ref{31r}) and
the related polaronic self-energy, we will use the static
dielectric function given in Eq. (\ref{31rz}).

First, we focus on $Z_F$, the spectral weight at the Fermi
wave-vector $k_F$, which is derived from the Green's function in
Eq. (\ref{136r}). Really, this renormalization factor plays the
role of the residue at the pole in a Fermi-liquid description
\cite{imada}. Thus, we can evaluate the renormalized electron
distribution function $n^{ren}_{{\bf k}}= \int_{- \infty}^{+
\infty} \frac{d E}{2 \pi} A( {\bf k}, E) n_F ( E )$. At zero
temperature, we recall that $ n^{ren}( \mu -\delta )-n^{ren}( \mu
+\delta )= Z_F $ (with $\delta$ infinitesimal energy), so that the
factor $Z_F$ determines the jump in the Fermi distribution
function \cite{alex}. Actually, the polaronic self-energy
introduces tiny differences between $Z_F$  and the renormalization
term $e^{-S}$ in Eq. (\ref{136r}) only in the low density regime,
since, with increasing the particle density, screening rapidly
reduces the effects due to polaron-polaron and polaron-phonon
interactions.

\begin{figure}[t]
\begin{center}
\includegraphics[angle=-90,scale=0.29]{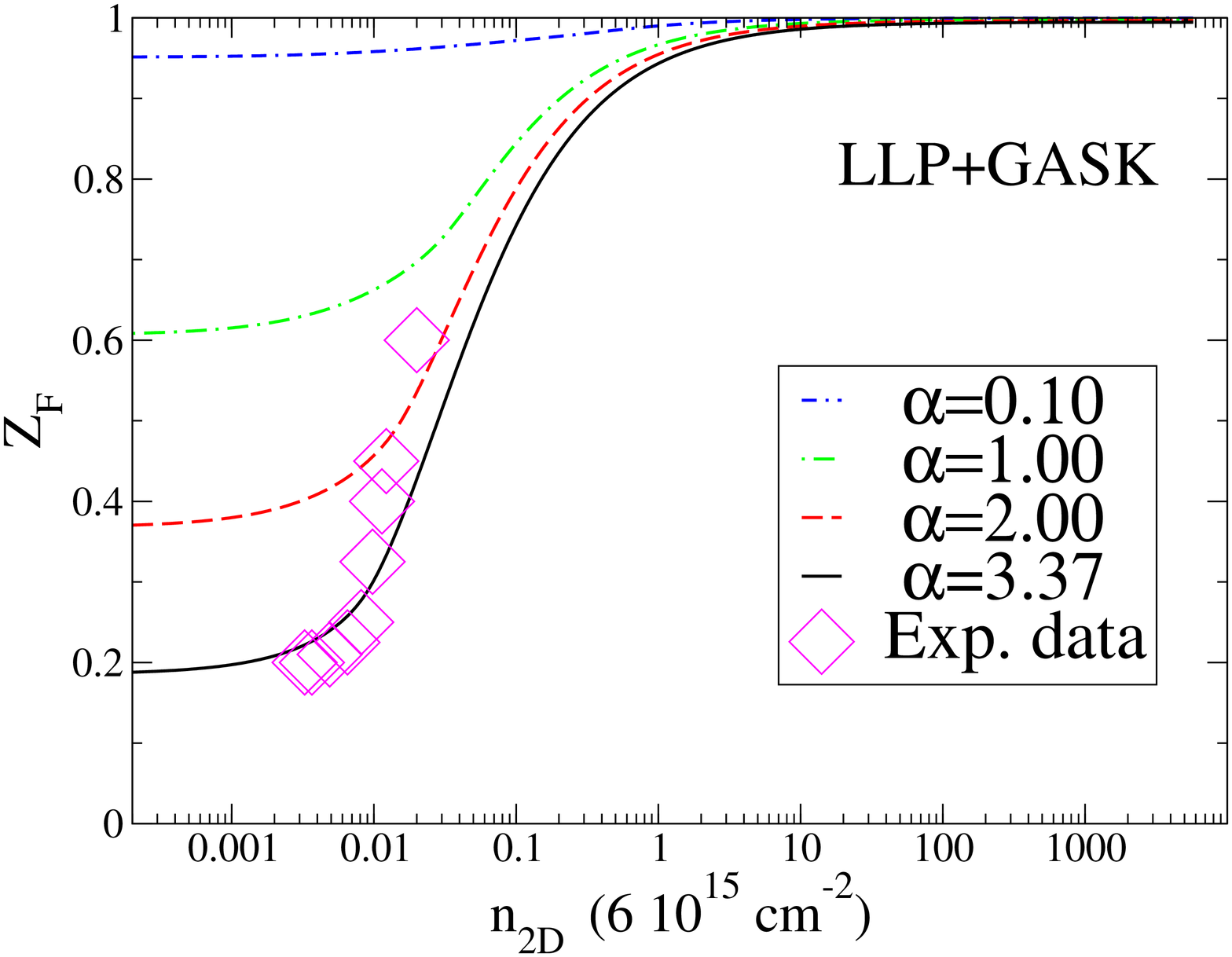} \\
\includegraphics[angle=-90,scale=0.29]{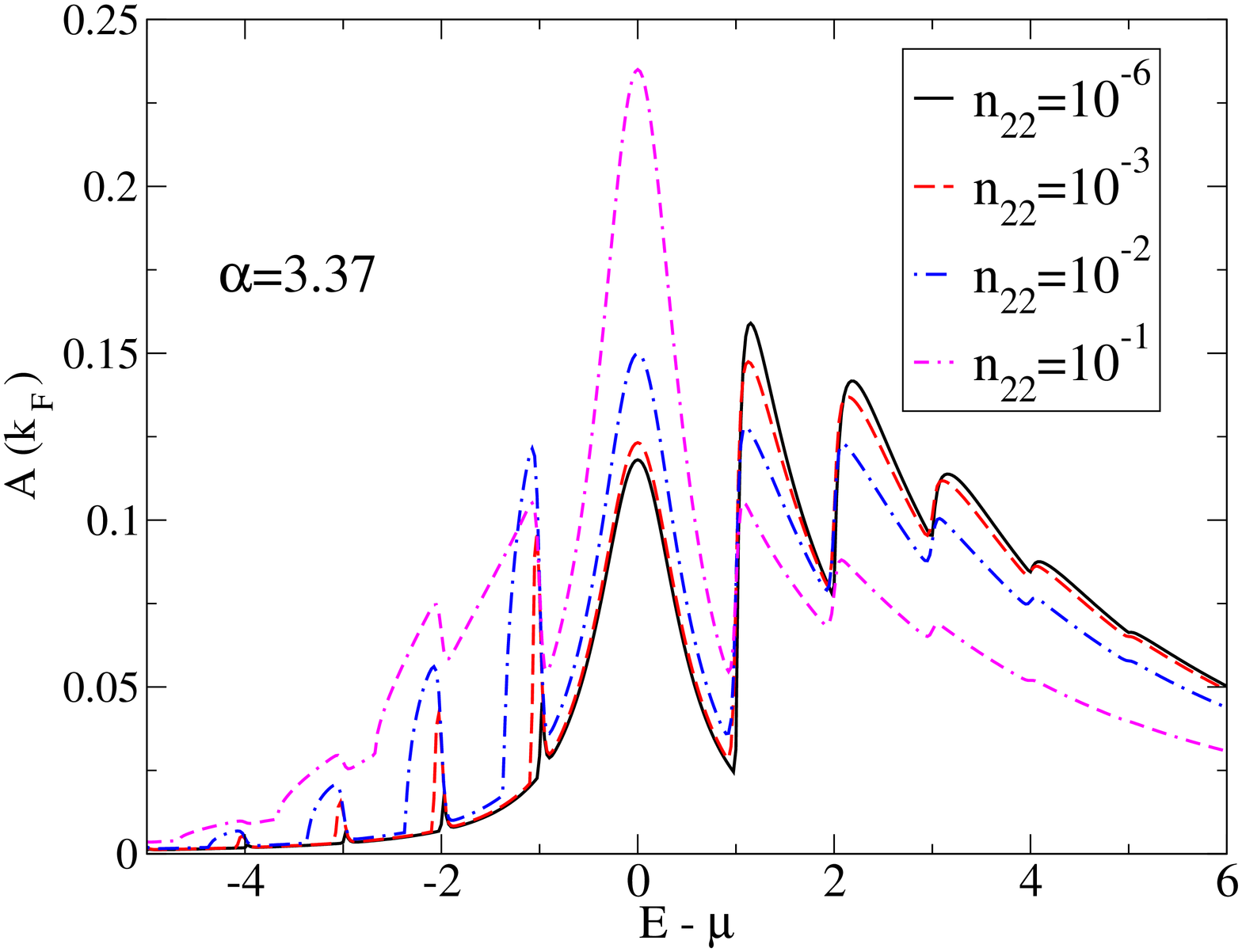}
\caption{ Upper Panel: Spectral weight at Fermi wave-vector $k_F$,
$Z_F$, as a function of the particle density $n_{22}$ for
different values of the electron-phonon constant $\alpha$ within
the many-body LLP approach with Gaskell-RPA treatment of
polaron-polaron interactions (LLP+GASK). Experimental data are
taken from Ref. \cite{devereaux}. Lower Panel: Spectral function
(in units of $1/\hbar \omega_{LO})$ at Fermi wave-vector $k_F$ as
a function of the energy (related to the chemical potential $\mu$,
both in units of $\hbar \omega_{LO}$) for different values of
particle density at $\alpha=3.37$ within the many-body
perturbative method upon the LLP approach. A width $\Gamma=0.5
\hbar \omega_0$ has been added as an imaginary part to the
polaronic self-energy.} \label{fig5}
\end{center}
\end{figure}

In the upper panel of Fig. \ref{fig5}, we plot $Z_F$ as a function
of the two-dimensional particle density $n_{2D}$ for different
values of the electron-phonon constant $\alpha$. We recall that
the two-dimensional $n_{2D}$ is obtained from the
three-dimensional density $n$ through the effective length $d=6$
nm. For any value of the coupling constant $\alpha$, $Z_F$ gets
enhanced with increasing the particle density. We stress that this
increase starts from around $n_{2D}=6 \cdot 10^{12}$ $cm^{-2}$,
and it is very rapid from $n_{2D}=6 \cdot 10^{13}$ $cm^{-2}$.
Indeed, at $n_{2D}=6 \cdot 10^{14}$ $cm^{-2}$, the spectral
weights for different values of $\alpha$ converge towards similar
values. Finally, at $n_{2D}=6 \cdot 10^{16}$ $cm^{-2}$, $Z_F$ is
close to unity for any value $\alpha$ indicating that the
screening of many-body interactions is almost complete. For these
values of the density, polaronic effects are very weak, therefore
the system presents a conventional metallic state.

%%%%%%%%%%%%%%%%%%%%%%%%%%%%%%
%%%%%%%%%%%%%%%%%%%%%%%%%%%%%}

It is important to note that these results are compatible with
photoemission experiments at the $(001)$ STO surface
\cite{devereaux}. Indeed, as shown in the upper panel of Fig.
\ref{fig5}, not only the low density value of $Z_F$ but also its
calculated behavior as a function of the particle density agree
with experimental data. In particular, with increasing the
particle density, the experimental enhancement of $Z_F$ looks a
little bit more marked than that predicted by theory. We notice
that the two-dimensional behavior is predicted by theory taking
into account the effective length $d$ of the electron gas. One
expects that screening effects beyond Hartree-Fock approximation
would be more pronounced in an actual two-dimensional calculation
providing a more rapid increase of the spectral weight $Z_F$.

The next step is to analyze the spectral function derived from the
Green's function in Eq. (\ref{136r}). In the lower panel of Fig.
\ref{fig5}, fixing $\alpha=3.37$, we report the spectral function
at the Fermi wave-vector $k_F$ as a function of the energy for
different densities considering both the hole ($E<\mu$) and the
particle sector ($E>\mu$). The peak-dip-hump line shape is
recovered not only for low densities, but also for high densities.
With increasing carrier concentration, a transfer of spectral
weight occurs towards the coherent peak by reducing the hump
consisting of phonon satellites due to the incoherent large
polaron dynamics. Moreover, this transfer towards the coherent
peak is accompanied by an increase of the spectral weight of the
hole sector with increasing density. We remark that experiments in
Ref. \cite{devereaux} show only the hole sector, which is very
carefully described by out theory. For example, as shown in the
lower panel of Fig. \ref{fig5}, in contrast with the behavior in
the particle sector, the coherent peak is higher than the first
phonon satellite in the hole sector for all the charge densities.
Actually, in the lower panel of Fig. \ref{fig5}, we have added
width $\Gamma=0.5 \hbar \omega_0$ as an imaginary part to the
polaronic self-energy in order to simulate the instrumental
resolution of photoemission spectra and to make a more realistic
comparison with experiments. Moreover, in agreement with
experiments \cite{devereaux}, we recognize three different density
regimes: the first one corresponding to quite low densities
($n_{22} \le 10^{-4}$), the second one to intermediate densities
($10^{-4}<n_{22}<10^{-1}$), the third one to high densities
($n_{22} \ge 10^{-1}$). In the final part of the paper, we analyze
these three regimes.

In the first regime at low densities, the spectral function is not
identical to that in the limit of single polaron discussed in a
previous subsection, since, as shown in the lower panel of Fig.
\ref{fig5} for $n_{22}=10^{-6}$, there is also the incoherent hole
contribution for energies below the chemical potential. Even if
the incoherent hole hump has less spectral weight than the
incoherent electron hump, the number of phonon satellites in the
two humps is very similar confirming that polaronic effects are
active both in the hole and particle channel. In this first regime
with low density, screening is not active, therefore, electrons
interact with phonons through a long range Fr\"ohlich coupling.

In the second regime of intermediate density
($10^{-4}<n_{22}<10^{-1}$), screening starts to reduce polaronic
effects whose spatial range decreases with increasing density.
Indeed, as shown in the lower panel of Fig. \ref{fig5} for
$n_{22}=10^{-3}$ and $n_{22}=10^{-2}$, the number of phonon
satellites is reduced transferring spectral weight to the coherent
peak. We stress that this coherent-incoherent crossover is quite
rapid: for $n_{22}=10^{-3}$ (corresponding to $n_{2D}$ of the
order of $10^{13}$ $cm^{-2}$), the satellite structure is not
dissimilar from that of lower densities, on the other hand, for
$n_{22}=10^{-2}$ (corresponding to $n_{2D}$ of the order of
$10^{14}$ $cm^{-2}$), only the first phonon satellite is marked,
the second one is strongly reduced, the third one has almost
completely disappeared. All these features are in excellent
agreement with tunneling and photoemission experiments probing the
polaronic liquid in STO-based systems \cite{devereaux,swartz}.

In the third regime at high densities ($n_{22} \ge 10^{-1}$),
screening becomes predominant causing the breakdown of the
polaronic state. As shown in the lower panel of Fig. \ref{fig5}
for $n_{22}=10^{-1}$, the weight of the coherent term is
prevalent. The system behaves as a metal with a short range
electron-phonon coupling. Indeed, in this regime, the mass ratio
at the Fermi wave-vector $k_F$, $\frac{m}{m_F^{*}}$, becomes very
similar to the spectral weight $Z_F$, confirming the short range
character of many-body interactions.

Summarizing, the variation of carrier concentration controls the
screening of many-body interactions affecting the spectral
properties of large polaron systems. From the comparison between
the lower panels of Figs. \ref{fig4} and \ref{fig5},  it emerges
that the role of density can be roughly understood as an effect
leading to the reduction of the electron-phonon coupling constant.
An estimate of this reduction as a function of the particle
density is not easy since it involves both polaron features and
many-body screening, which, in our work, are intimately linked.

\section{Conclusions and discussions}

In this paper, we have discussed ground state and spectral
properties of the Fr\"ohlich model as a function of the particle
density focusing on the intermediate electron-phonon coupling
regime at zero temperature. We have introduced a new variational
approach exploring a huge range of particle densities. The
formation of the large polaron and the role of screening turn out
to play a crucial role in understanding the spectral properties of
STO-based systems with varying the carrier concentration. In the
case of a single polaron, the peak-dip-hump line shape is in good
agreement with the spectral function obtained by numerical
approaches and with experimental spectra of STO-based systems in
the low density limit. In addition to the low density regime, we
have identified other two relevant density ranges, the
intermediate and high density ones. While for high densities the
system shows a conventional metallic phase, for intermediate
densities, a rapid crossover takes place from incoherent to
coherent large polaron dynamics with increasing carrier density
finding very good agreement with experimental spectra in STO-based
systems.

%%%%%%%%%%%%%%%%%%%%%%%%%%%%%%%%%%%%%%%%%%%%%%%%%%%%%%%%%%%%%%%%%%%
\begin{figure*} [t!]
\includegraphics[scale=0.2]{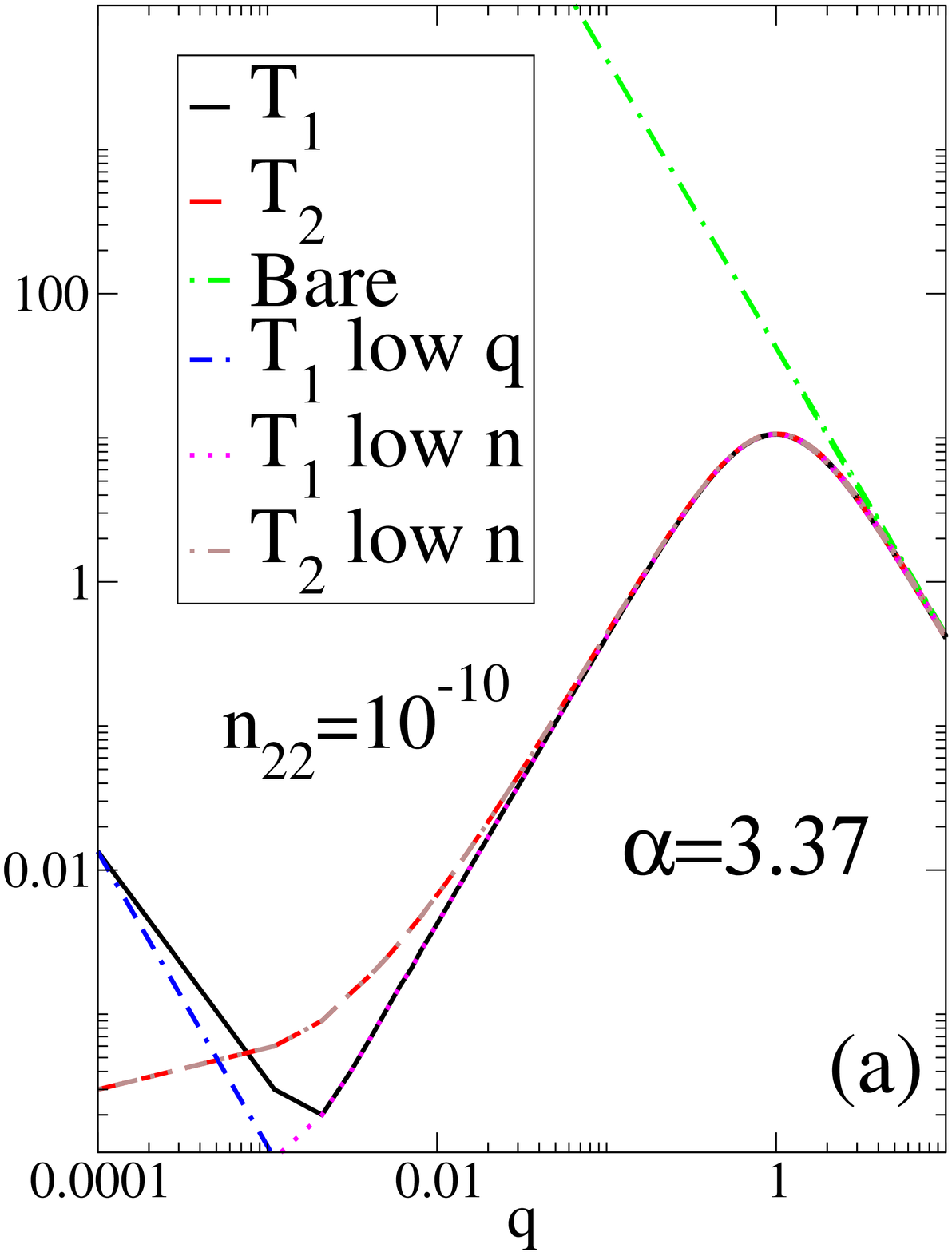}
\includegraphics[scale=0.2]{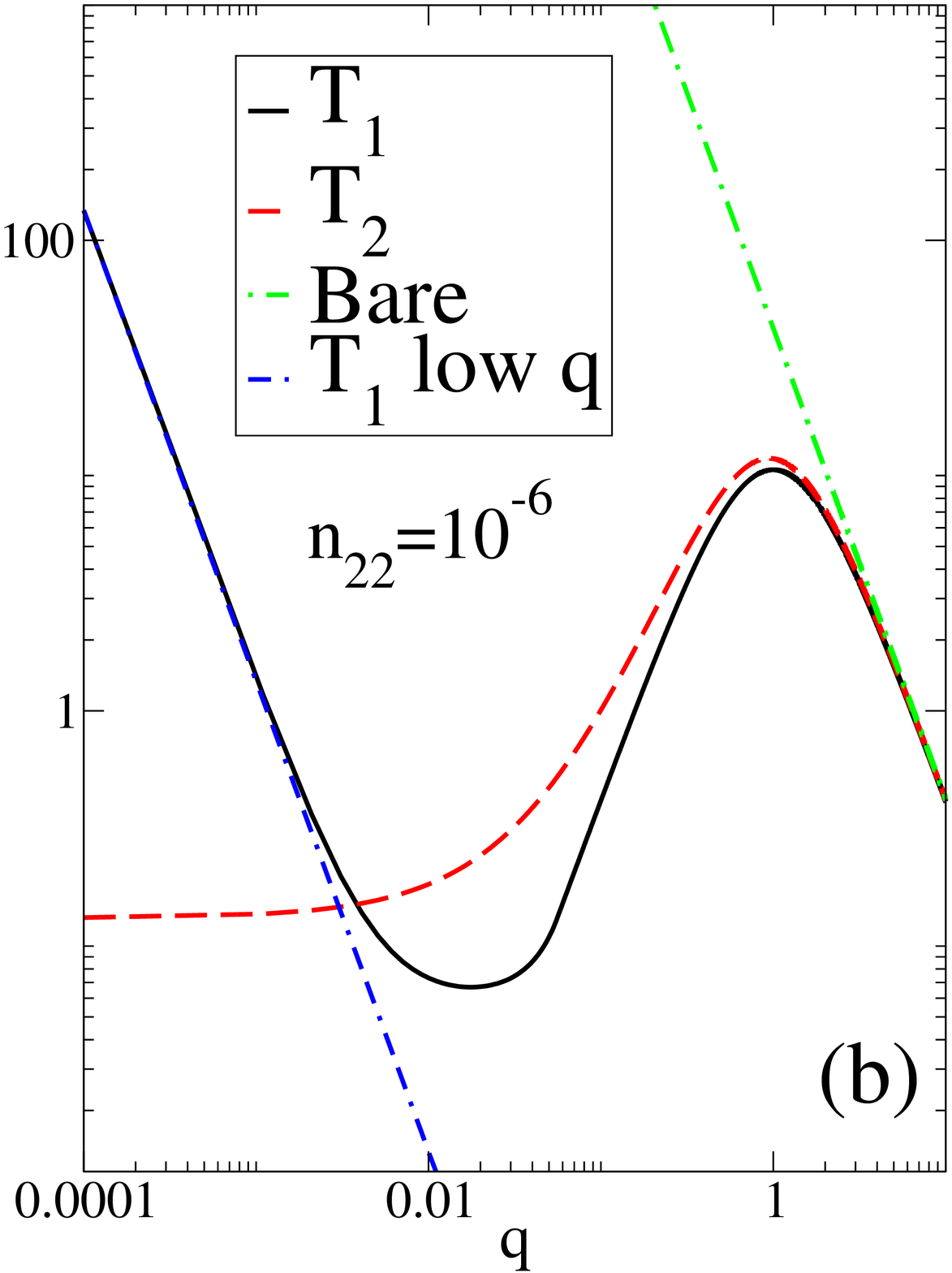}
\includegraphics[scale=0.2]{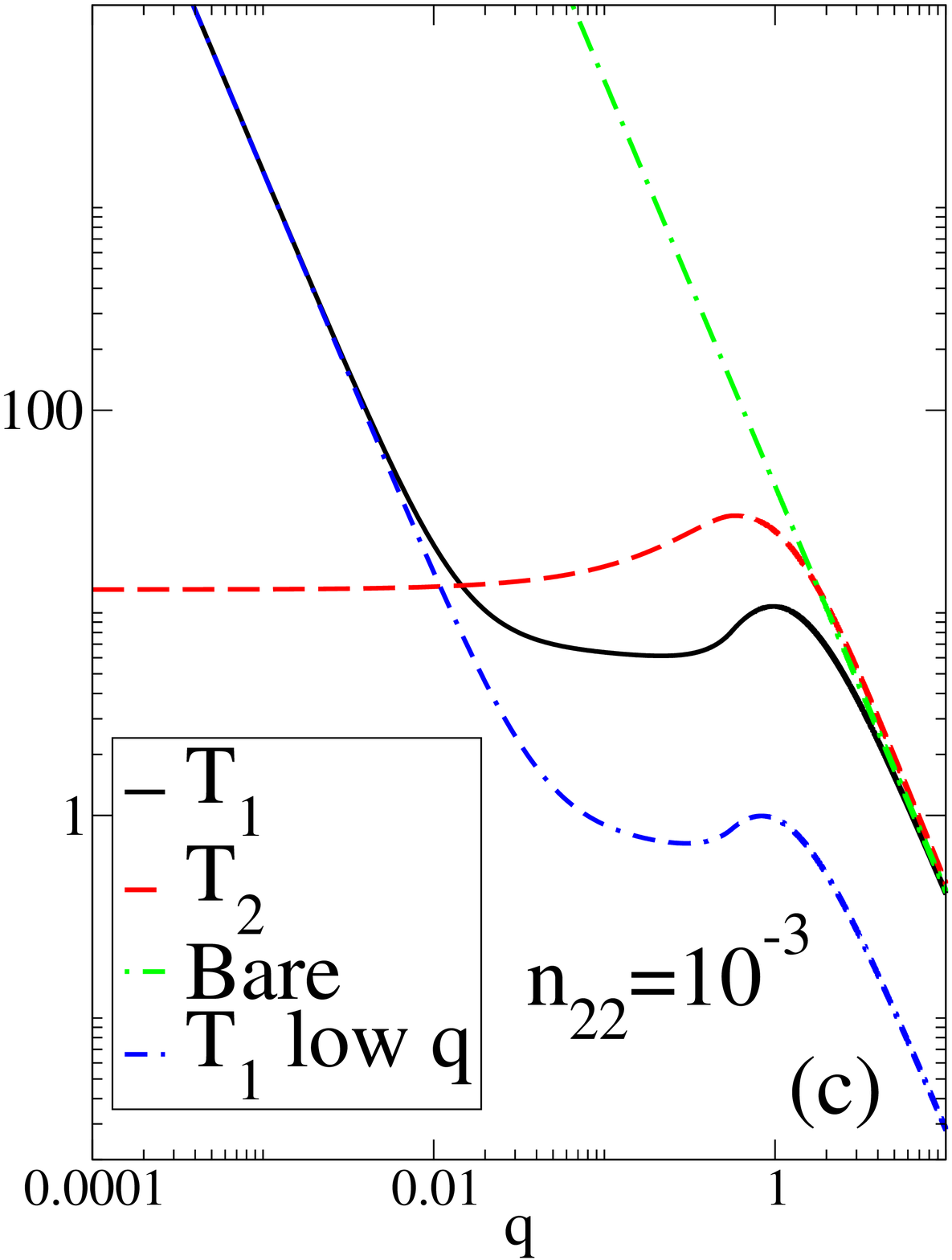}
\includegraphics[scale=0.2]{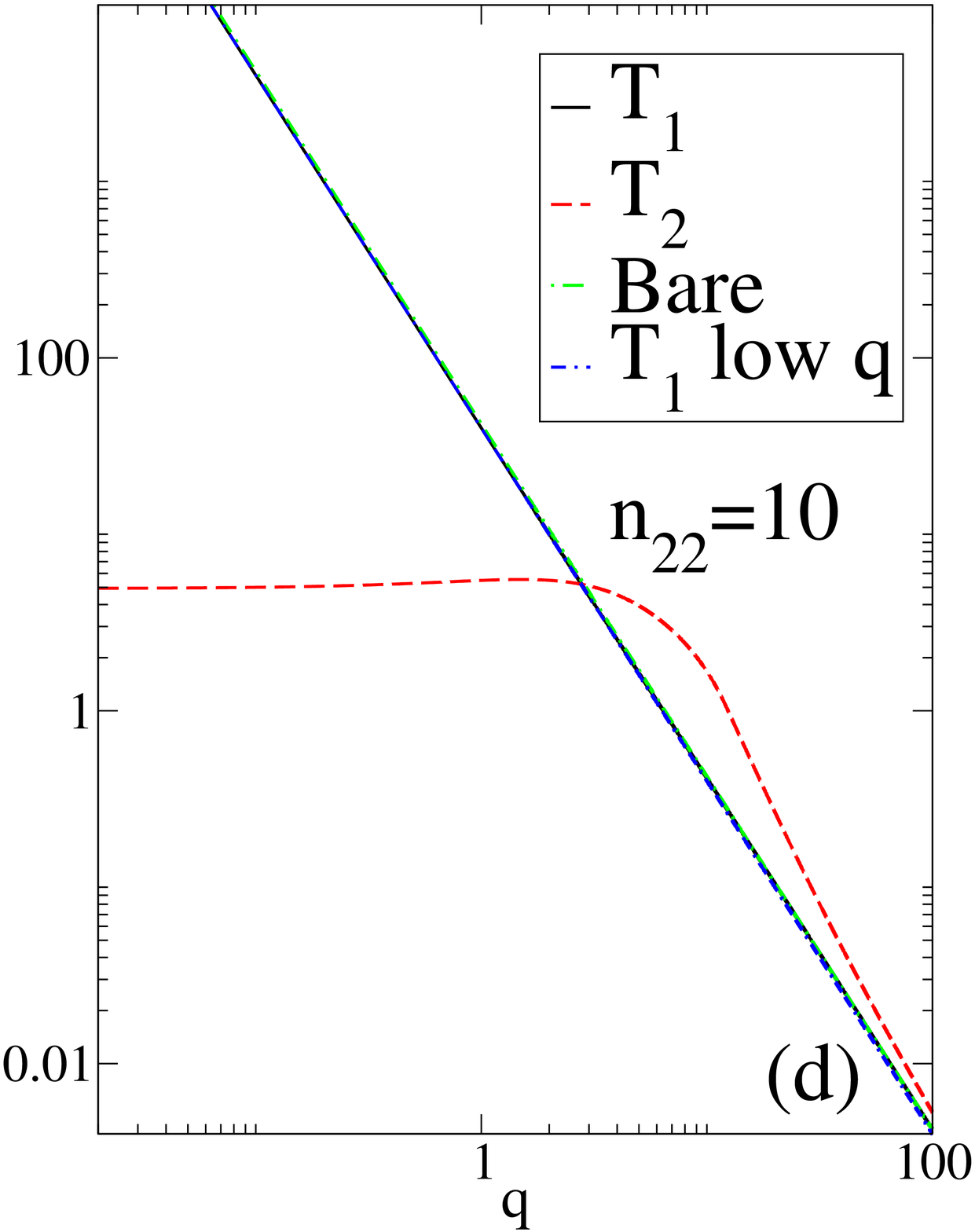}
\caption{Polaron-phonon couplings (in units of $ (\hbar
\omega_{LO})^2 R^3_P$) as a function of the wave-vector $q$ (in
units of $1/R_P$) at electron-phonon coupling constant
$\alpha=3.37$ for $n_{22}=10^{-10}$ (Panel (a)), $n_{22}=10^{-6}$
(Panel (b)), $n_{22}=10^{-3}$ (Panel (c)), $n_{22}=10$ (Panel
(d)). $T_1$ $low$ $q$ stands for the low $q$ expansion of the
coupling $T_1$, $T_1$ $low$ $n$ for the low density expansion of
the coupling $T_1$, $T_2$ $low$ $n$ for the low density expansion
of the coupling $T_2$, $Bare$ for the bare coupling given by the
square of $M_q$ defined in Eq. (\ref{2rbis}). } \label{fig3}
\end{figure*}
%%%%%%%%%%%%%%%%%%%%%%%%%%%%%%%%%%%%%%%%%%%%%%%%%%%%%%%%%%%%%%%%%%%

In this work, we have ascribed all the electron-phonon coupling to
a single longitudinal optical mode with frequency $\omega_{LO}$
which is mostly coupled to charge carriers. This mode has quite a
high frequency ($\hbar \omega_{LO} \simeq 100$ meV), therefore we
have largely analyzed the antiadiabatic regime relative to this
mode (Fermi energy $E_F$ such that $E_F < \hbar \omega_{LO}$)
Additional low frequency optical modes are present, however, they
are much more weakly coupled to the electrons
\cite{strocov,swartz}. Tiny spectral features due to these modes
can be recognized in tunneling experiments \cite{swartz}, but are
not visible in photoemission data \cite{devereaux}. Indeed, due to
instrumental resolution of photoemission experiments, the main
peak shown in the experimental data at the Fermi energy is quite
large. We have estimated a width of the order of $\Gamma=0.5 \hbar
\omega_0$. Therefore, in photoemission experiments, the main peak
at the Fermi energy could include not only the coherent peak but
also the first satellites due to low frequency optical modes. This
is the reason why, in the literature, the modeling of spectral
properties has been done in the perturbative electron-phonon
coupling regime considering only the most coupled high frequency
mode \cite{devereaux}. We point out that additional phonon modes
can be included into the theoretical model since this involves a
simple generalization of our approach.

In this work, we have focused on the spectral properties of the
normal state at zero temperature. The next step could be the
analysis of superconducting states \cite{lin1,lin2} where the
coupling to longitudinal optical phonons plays a non negligible
role \cite{mannhart,gorkov}. Finally, another interesting aspect
could be related to the role of electron-phonon coupling on the
temperature behavior of spectral and transport properties
\cite{cancellieri,Zhou,perroni4}, for example of the
thermoelectric Seebeck effect \cite{pallecchi,perroni3}.

\section*{Acknowledgments}
C.A.P. acknowledges support by the project QUANTOX (QUANtum
Technologies with 2D-OXides) of QuantERA-NET Cofund in Quantum
Technologies, implemented within the EU-H2020 Programme, and the
project TOPSPIN (Two-dimensional Oxides Platform for
SPINorbitronics nanotechnology) funded by the MIUR-PRIN Bando 2017
- grant 20177SL7HC.

\appendix

\section{Polaron-phonon couplings}

In this Appendix, we discuss  the polaron-phonon couplings of the
transformed Hamiltonian $\tilde{H}$ in Eq. (\ref{6r}), which, in
addition to the polaron-polaron potential $V_q^{eff}$, are
relevant for the evaluation of the spectral properties presented
in the main text.

We start considering the first coupling $T_{1,q}$, defined as
\begin{equation}
T_{1,q}=\left(M_q^{eff}\right)^2=M_q^2 \frac{\left[ \frac{\hbar^2
q^2}{2 m S_q^{eff}} \right]^2}{\left[\hbar \omega_{LO} +
\frac{\hbar^2 q^2}{2 m S_q^{eff}} \right]^2}, \label{27r_ter4}
\end{equation}
where we have used the expression of $M_q^{eff}$ in Eq.
(\ref{11r}). Within the Gaskell approach, for small values of the
wave-vector $q$,
\begin{equation}
T_{1,q} \simeq M_q^2 \frac{\left[ \omega_{PP}
\right]^2}{\left[\omega_{LO} + \omega_{PP}  \right]^2}.
\label{27r_ter2}
\end{equation}
For low particle densities such that $\omega_{PP} \ll
\omega_{LO}$, with $\omega_{PP}$ the polaron plasmon at zero
wave-vector given in Eq. (\ref{24rr}), in the limit of small $q$,
$T_{1,q}$ is quite small. Otherwise, for large particle densities
such that $\omega_{PP} \gg \omega_{LO}$, in the limit of small
$q$, $T_{1,q}$ tends to the bare coupling $M_q^{2}$. Finally, for
large values of $q$, as expected, $T_{1,q} \simeq M_q^2$.

Then, we study the second coupling $T_{2,q}$, defined as
\begin{equation}
T_{2,q}=\left[ N_{{\bf k}_1, {\bf k}_1+{\bf q}} \right ]^2=
 \left( \frac{\hbar^2 f_q}{2m} \right)^2 \left[ {\bf q} \cdot ({\bf q}+2 {\bf
k}_1 \right]^2, \label{27r_ter3}
\end{equation}
where we have considered the polaron-phonon matrix element in Eq.
(\ref{12rbis}), with ${\bf k}_1=k_F {\hat q}$. Therefore, for this
coupling, we consider the most relevant contribution, that is that
at the Fermi wave-vector $k_F$ in a direction given by the versor
$ {\hat q}$ of the wave-vector ${\bf q}$. In analogy with
$T_{1,q}$, $T_{2,q}$ tends towards the bare coupling $M_q^{2}$ for
large values of $q$.

In Fig. \ref{fig3} we plot the first coupling $T_{1,q}$ and the
second coupling $T_{2,q}$ as a function of the modulus $q$ of the
wave-vector ${\bf q}$ for different particle densities. The panel
(a) corresponds to the lowest particle density $n_{22}=10^{-10}$.
We notice that, for this density, $T_{1,q}$ and $T_{2,q}$ are
almost identical for a large range of values of $q$. They differ
only for small values of $q$, where $T_{1,q}$ increases with
decreasing $q$ recovering the limit for small $q$. On the other
hand, $T_{2,q}$ is always coincident with the limit for small
density $n$.

The panel (b) of Fig. \ref{fig3} corresponds to the low particle
density $n_{22}=10^{-6}$. The couplings $T_{1,q}$ and $T_{2,q}$
coincide for large values of $q$. For intermediate values of $q$,
$T_{1,q}$ is smaller than $T_{2,q}$, while, for small values of
$q$, one gets the opposite. In fact,  $T_{1,q}$ shows a crossover
from the small $q$ limit to the bare coupling with increasing the
values of $q$.

In panel (c) of Fig. \ref{fig3} we plot the couplings $T_{1,q}$
and $T_{2,q}$ for the particle density $n_{22}=10^{-3}$. They
differ in a large range of values of $q$. The coupling $T_{1,q}$
shows a narrower crossover from the small $q$ limit to the bare
coupling. Finally, the panel (d) of Fig. \ref{fig3} shows the
couplings $T_{1,q}$ and $T_{2,q}$ for the particle density
$n_{22}=10$. As expected, the coupling $T_{1,q}$ always coincides
with the bare coupling, while the coupling $T_{2,q}$ is negligible
in comparison with $T_{1,q}$ for a large range of values of $q$.
As discussed in the main text, for high values of density,
screening of the polaron-phonon vertex is fundamental to properly
calculate the spectral properties.

%%%%%%%%%%%%%%%%%%%%%%%%%%%%%%
\begin{figure}[t]
\begin{center}
\includegraphics[angle=0,width=8.5cm]{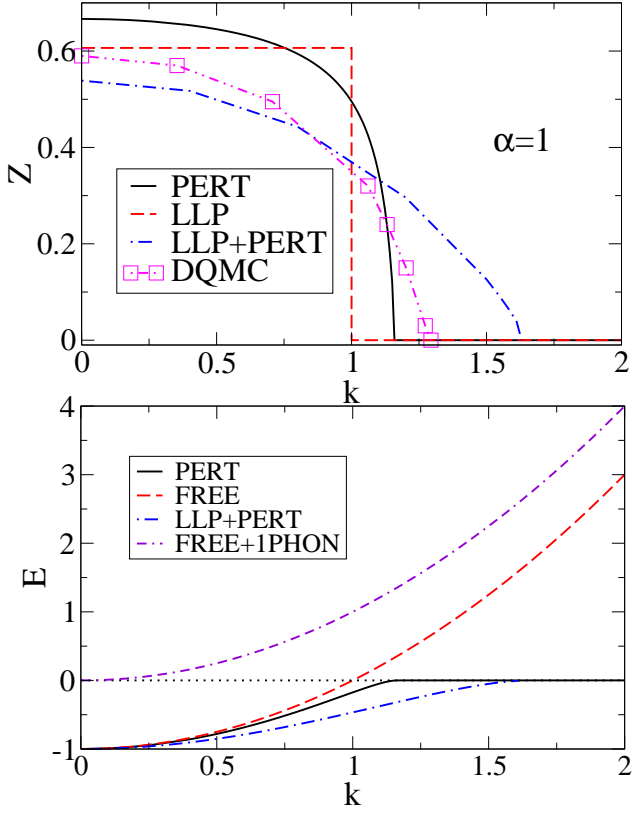}
\caption{ Upper Panel: Single fermion spectral weight as a
function of the wave-vector $k$ (in units of $1/R_P$) for
different approaches: $PERT$ stands for perturbative approach,
$LLP$ for Lee-Low-Pines method, $LLP+PERT$ for perturbative
corrections upon LLP method, $DQMC$ for Diagrammatic Quantum
Monte-Carlo data from Ref. \cite{andrei1}. Lower Panel: Single
fermion quasi-particle energy (in units of $\hbar \omega_{LO})$ as
a function of the wave-vector $k$ (in units of $1/R_P$) for
different approaches: $PERT$ stands for perturbative approach,
$LLP+PERT$ for perturbative corrections upon LLP method, $FREE$
for the bare electronic dispersion (shifted by the ground state
energy $E_0$), $FREE+1PHON$ for the same with a shift of $\hbar
\omega_{LO}$. In both panels, electron-phonon coupling constant
$\alpha=1$. } \label{fig1A}
\end{center}
\end{figure}
%%%%%%%%%%%%%%%%%%%%%%%%%%%%%}

\section{Additional results on polaron spectral properties}

In this Appendix, we provide some details concerning the
dispersion of the quasi-particles as a function of the
wave-vector. In particular, we discuss the evaluation of the
polaronic self-energy in the case of wave-vectors different from
Fermi wave-vector $k_F$.

In this paper, a perturbative approach is made on top of the LLP
scheme, which, however, already provides quite accurate polaronic
energies. Therefore, in our perturbative approach, we fix $\Re
\Sigma^{ret}_{pol} (k_F,\omega=0)=0$ \cite{giulio}, therefore the
chemical potential is $\mu=\frac{\hbar^2 k_F^2}{2m}+ \eta$, where
$\eta$ is the polaronic band shift given in Eq. (\ref{8r}). In the
limit of single polaron, one gets the correct ground state energy
$E_0=\eta=-\alpha \hbar \omega_{LO}$ within the intermediate
electron-phonon coupling regime.

In order to investigate the behavior of spectral properties as a
function of the wave-vector $k$, we analyze the single polaron
case. We plot the spectral weight $Z_k$ as a function of the
wave-vector $k$ in the upper panel of Fig. \ref{fig1A} comparing
different approaches at $\alpha=1$. We notice the rapid decrease
of the spectral weight with increasing $k$. The LLP scheme is not
able to describe this behavior, while the lowest order
perturbation theory on top of the LLP approach ($LLP+PERT$ in
figure) is able to interpolate the DQMC numerical data. Clearly,
the comparison of the approach with DQMC results could improve if
one would include further corrections in the building up of the
self-energy. For example, one possibility would be to make a
self-consistent calculation between polaronic Green's function and
self-energy.

In addition to the spectral weight, we have derived the polaron
dispersion. We plot the quasi-particle energy as a function of the
wave-vector $k$ in the lower panel of Fig. \ref{fig1A} comparing
different approaches at $\alpha=1$. There is a flattening of the
polaron dispersion when the spectral weight goes to zero. This
occurs when the difference between the energy at finite $k$ and
that at $k=0$ is equal to the phonon energy $\hbar \omega_{LO}$
\cite{mahan}. Finally, we consider the effective mass analyzing
the polaron dispersion for small wave-vectors. For $\alpha=1$, the
approach perturbative in the electron-phonon coupling provides the
estimate $m_0^{*} \simeq 1.13 m $, by using Eq. (\ref{110r}). As
shown in the lower panel of Fig. \ref{fig1A}, the dispersion
calculated within the perturbation theory upon the LLP scheme has
a smaller curvature. Indeed, the effective mass ratio is a little
bit higher: $m_0^{*} \simeq 1.2 m $.


\begin{thebibliography}{100}
\bibitem{notaperturbazione}
In this paper, we use a pertubative approach for the
electron-phonon interaction which does not show any anomalous
behavior with increasing the coupling constant $\alpha$.
\end{thebibliography}
\end{document}